\documentclass[longauth]{aa}  

\usepackage[utf8]{inputenc}
\usepackage{graphicx}
\usepackage{multirow}
\usepackage{amsmath}
\usepackage{txfonts}
\usepackage{natbib}
\bibpunct{(}{)}{;}{a}{}{,} 
\usepackage{hyperref}
\hypersetup{
    colorlinks=true,
    citecolor=blue,
    filecolor=magenta
}
\begin{document}

   \title{COSMOS-Web galaxy groups: \\ Evolution of red sequence and quiescent galaxy fraction}

\author{Greta Toni\inst{1,2,3}\thanks{\email{greta.toni4@unibo.it}}
\and Matteo Maturi\inst{3,4}
\and Gianluca Castignani\inst{2}
\and Lauro Moscardini\inst{1,2,5}
\and Ghassem Gozaliasl\inst{6,7}
\and Alexis Finoguenov\inst{7}
\and Sina Taamoli\inst{8}
\and Fabrizio Gentile\inst{9,2}
\and Hollis B. Akins\inst{10}
\and Rafael C. Arango-Toro\inst{11}
\and Caitlin M. Casey\inst{12,13}
\and Nicole E. Drakos\inst{14}
\and Andreas L. Faisst\inst{15}
\and Carter Flayhart\inst{16}
\and Maximilien Franco\inst{9}
\and Ali Hadi\inst{8}
\and Aryana Haghjoo\inst{8}
\and Santosh Harish\inst{16}
\and Hossein Hatamnia\inst{8}
\and Olivier Ilbert\inst{11}
\and Shuowen Jin\inst{13,17}
\and Jeyhan S. Kartaltepe\inst{16}
\and Ali Ahmad Khostovan\inst{16,18}
\and Anton M. Koekemoer\inst{19}
\and Gavin Leroy\inst{20}
\and Georgios E. Magdis\inst{13,17,21}
\and Henry Joy McCracken\inst{22}
\and Jed McKinney\inst{10}
\and Louise Paquereau\inst{22}
\and Jason Rhodes\inst{23}
\and R. Michael Rich\inst{24}
\and Brant E. Robertson\inst{25}
\and Rasha M. Samir\inst{26}
\and Diana Scognamiglio\inst{23}
\and Samaneh Shamyati\inst{8}
\and Marko Shuntov\inst{13,21,27}
\and Jorge A. Zavala\inst{28}
}

\institute{University of Bologna, Department of Physics and Astronomy “Augusto Righi” (DIFA), Via Gobetti 93/2, I-40129 Bologna, Italy \and
INAF – Osservatorio di Astrofisica e Scienza dello Spazio, Via Gobetti 93/3, I-40129 Bologna, Italy \and
Zentrum f\"ur Astronomie, Universit\"at Heidelberg, Philosophenweg 12, D-69120 Heidelberg, Germany \and
Institut f\"ur Theoretische Physik, Universit\"at Heidelberg, Philosophenweg 16, D-69120 Heidelberg, Germany \and
INFN – Sezione di Bologna, Viale Berti Pichat 6/2, I-40127 Bologna, Italy \and
Department of Computer Science, Aalto University, P.O. Box 15400, FI-00076 Espoo, Finland \and
Department of Physics, University of Helsinki, P.O. Box 64, FI-00014 Helsinki, Finland \and
Department of Physics and Astronomy, University of California, Riverside, 900 University Avenue, Riverside, CA 92521, USA \and
Université Paris-Saclay, Université Paris Cité, CEA, CNRS, AIM, 91191 Gif-sur-Yvette, France \and
Department of Astronomy, The University of Texas at Austin, 2515 Speedway Blvd Stop C1400, Austin, TX 78712, USA \and
Aix Marseille Univ, CNRS, CNES, LAM, Marseille, France \and
Department of Physics, University of California, Santa Barbara, Santa Barbara, CA 93106, USA \and
Cosmic Dawn Center (DAWN), Denmark \and
Department of Physics and Astronomy, University of Hawaii, Hilo, 200 W Kawili St, Hilo, HI 96720, USA \and
Caltech/IPAC, MS 314-6, 1200 E. California Blvd., Pasadena, CA 91125, USA \and
Laboratory for Multiwavelength Astrophysics, School of Physics and Astronomy, Rochester Institute of Technology, 84 Lomb Memorial Drive, Rochester, NY 14623, USA \and
DTU Space, Technical University of Denmark, Elektrovej 327, 2800 Kgs. Lyngby, Denmark \and
Department of Physics and Astronomy, University of Kentucky, 505 Rose Street, Lexington, KY 40506, USA \and
Space Telescope Science Institute, 3700 San Martin Drive, Baltimore, MD 21218, USA \and
Institute for Computational Cosmology, Department of Physics, Durham University, South Road, Durham DH1 3LE, United Kingdom \and
Niels Bohr Institute, University of Copenhagen, Jagtvej 128, DK-2200 Copenhagen, Denmark \and
Institut d’Astrophysique de Paris, UMR 7095, CNRS, and Sorbonne Université, 98 bis boulevard Arago, F-75014 Paris, France \and
Jet Propulsion Laboratory, California Institute of Technology, 4800 Oak Grove Drive, Pasadena, CA 91001, USA \and
Department of Physics and Astronomy, UCLA, PAB 430 Portola Plaza, Box 951547, Los Angeles, CA 90095-1547, USA \and
Department of Astronomy and Astrophysics, University of California, Santa Cruz, 1156 High Street, Santa Cruz, CA 95064, USA \and
National Research Institute of Astronomy and Geophysics (NRIAG), Cairo, Egypt \and
University of Geneva, 24 rue du Général-Dufour, 1211 Genève 4, Switzerland \and
University of Massachusetts Amherst, 710 North Pleasant Street, Amherst, MA 01003-9305, USA
}

  \abstract
   {}
   {We investigate the redshift evolution and group richness dependence of the quiescent galaxy fraction and red sequence (RS) parameters in COSMOS galaxy groups, spanning a wide redshift range, from $z=0$ to $z=3.7$.}
   {We analyzed the deep and well-characterized sample of groups recently detected with the AMICO algorithm in the COSMOS(-Web) field. Our study of the quiescent galaxy population is based on a machine-learning classification tool based on rest-frame magnitudes. The algorithm learns from several traditional methods to estimate the probability of a galaxy being quiescent, achieving high precision and recall. Starting from this classification, we computed quiescent galaxy fractions within groups via two methods: one based on the membership probabilities provided by AMICO, which rely on an analytical model, and another using a model-independent technique.
   We then detected the RS by estimating the ridgeline position using probability-weighted photometric data, followed by $\sigma$ clipping to remove outliers. This analysis was performed using both rest-frame magnitudes and observer-frame magnitudes with rest-frame matching. We compared the results from both approaches and investigated the redshift and richness dependence of the RS parameters.}
   {We found that the quiescent galaxy population in groups builds up steadily from $z = 1.5-2$ across all richnesses, with faster and earlier growth in the richest groups. The first galaxies settle onto the RS ridgeline by $z \sim 2$, consistent with current evolutionary scenarios. Notably, we reported a rare protocluster core hosting quiescent galaxies at $z = 3.4$, potentially one of the most distant early RSs observed. Extending our study to X-ray properties, we found that X-ray faint groups have, on average, lower quiescent fractions than X-ray bright ones, likely reflecting their typical location in filaments where pre-processing is lower. Leveraging the broad wavelength coverage of COSMOS2025, we traced RS evolution using observed and rest-frame colors over $\sim 12$ Gyr, finding no significant trends in either the slope or the scatter of the ridgeline.}
   {}


   \maketitle

\section{Introduction}

Dense environments typically host galaxy populations characterized by lower star formation rates (SFRs) and redder stellar populations compared to lower-density environments, at least up to $z \sim 1$. Compared to those in the field, these galaxies are believed to undergo substantial transformation, not only in the SFR \citep[e.g.,][]{scoville_evolution_2013,darvish_effects_2016,taamoli_cosmos2020_2024}, but also in morphology \citep[e.g.,][]{capak_effects_2007,bamford_galaxy_2009}, gas content, and metallicity \citep[e.g.,][]{catinella_galex_2013}. This occurs through a variety of processes such as interactions \citep[][]{hausman_galactic_1978}, galaxy harrassment \citep{moore_galaxy_1996}, and ram-pressure stripping (\citealt{gunn_infall_1972}; see e.g., \citealt{boselli_environmental_2006}, for a review).  This environmental segregation is reflected in the
well-known SFR (or morphology)-density relation \citep[e.g.,][]{dressler_galaxy_1980,balogh_dependence_1998}. The way galaxies transform is generally explained as the result of a strongly intertwined impact of intrinsic properties, such as stellar mass (nature) and factors related to the different environments galaxies experience in their lifetimes (nurture) \citep[e.g.,][]{de_lucia_environmental_2012}. The disentanglement of mass- and environment-driven quenching remains debated. Building on the pioneering work of \cite{peng_mass_2010} establishing this two-channel framework, recent studies \citep[e.g.,][]{chartab_large-scale_2020,taamoli_cosmos2020_2024,zheng_photometric_2025, hatamnia_large-scale_2025} have expanded this view, showing that mass quenching is a strong driver across all cosmic times especially on very massive galaxies, while environmental quenching dominates the quenching of galaxies in dense environments at later epochs and especially for satellites, generating a popualtion of quiescent, red, passively evolving galaxies \citep[e.g.,][]{wetzel_galaxy_2012,ziparo2014,darvish_effects_2016,Kawinwanichakij2017}. This population, hosted in dense environments such as clusters and groups, is known to occupy a tight grouping in the color-magnitude diagram (CMD) of the galaxy system, known as the red sequence (RS). The study of this prominent feature in the CMDs of galaxy clusters and groups is crucial for tracing the formation and evolution of quiescent galaxies across cosmic time. 

The RS is thought to reflect the combined effects of early star formation histories, chemical enrichment, and environmentally driven quenching. These processes can be probed by studying the parameters of the RS ridgeline. The slope of the RS primarily encodes the mass–metallicity relation: more massive galaxies tend to be more metal-rich, and therefore redder, producing a negative color–magnitude slope \citep[e.g.,][]{kodama_origin_1997,gallazzi_ages_2006,de_lucia_build-up_2007,stott_evolution_2009}. The zero-point (the intercept or mean color at a fixed magnitude or stellar mass) is sensitive to the average stellar age and metallicity of the population, and its evolution with redshift constrains the formation epoch and passive aging of quiescent galaxies \citep[e.g.,][]{stanford_evolution_1998,gallazzi_ages_2006,mei_evolution_2009}. The scatter around the RS reflects the diversity of star formation histories, including residual or recent quenching, metallicity spread, and photometric uncertainties as well; a small intrinsic scatter indicates that most RS galaxies formed the bulk of their stars early and rapidly \citep[e.g.,][]{bower_precision_1992, blakeslee_clusters_2006,romeo_study_2015}. Taken together, the slope, zero-point, and scatter provide powerful diagnostics of the assembly history of quiescent galaxies and the timescales over which the RS is built up across cosmic time. 

The RS and quiescent fraction have been extensively studied in massive clusters \citep[e.g.,][]{mei_evolution_2009, stott_evolution_2009, hennig_galaxy_2017}. However, characterizing them in galaxy groups -- which dominate the halo mass function and host the bulk of galaxies in the Universe -- can provide deeper insights into the different quenching processes and galaxy formation scenarios. 
Over the past years, the presence of a well-defined RS in clusters and groups has also been used as a tracer to identify these systems and create robust catalogs up to $z\sim 1$ \citep[e.g.,][]{gladders_new_2000, rykoff_redmapper_2014,rykoff_redmapper_2016}. While quiescent fractions and RS properties have been well characterized in the local Universe and up to intermediate redshifts \citep[$z\sim 1$; e.g.,][]{menci_red_2008,rudnick_rest-frame_2009,fritz_vimos_2014}, a better understanding of how the first galaxies quenched and built up the RS can be addressed by studying samples of groups and clusters at higher redshifts, up to the earlier stages of cluster formation itself \citep[$z > 1.5-2$; e.g.,][]{kravtsov_formation_2012,shimakawa_mahalo_2018}. This epoch reveals a diversity in galaxy populations, with systems showing enhanced star formation, active galactic nucleus (AGN) activity, and merging activity \citep[e.g.,][]{brodwin_era_2013,alberts_star_2016,wang_discovery_2016}, and others (sometimes even coexisting) showing an early-phase RS and high red fractions, especially in the central regions of galaxy overdensities \citep[e.g.,][]{andreon_red_2011,spitler_first_2012,strazzullo_galaxy_2013,strazzullo_red_2016,zavala_gas_2019}. 

The James Webb Space Telescope (JWST) has recently demonstrated its unique power to detect quiescent galaxies at high redshift and to probe their environments \citep[e.g.,][]{jin_cosmic_2024,ito_merging_2025,de_graaff_efficient_2025}.
In this work, we aim to characterize quiescent fractions and RS in groups and low-mass clusters, across $\sim$ 12 Gyrs of cosmic history. To do so, we leveraged our recent work \citep{toni_cosmos-web_2025} in building the largest group catalog based on deep JWST observations to date, spanning $z=0.08$ to $z=3.7$, over the COSMOS-Web field \citep{casey_cosmos-web_2023}. In this catalog, galaxy groups have been detected with the Adaptive Matched Identifier of Clustered Objects \citep[AMICO;][]{bellagamba_amico_2018,maturi_amico_2019}.
The AMICO algorithm, officially selected for cluster detection in Euclid \citep{euclid_collaboration_adam_euclid_2019, euclid_collaboration_bhargava_euclid_2025}, is based on optimal linear matched filtering, and has been proven to be a flexible tool for identifying groups and even protocluster cores (or overdensity peaks) up to $ z\sim 4$ and down to a few $10^{12}$ M$_\odot$ \citep{toni_amico-cosmos_2024,toni_cosmos-web_2025}. What is particularly interesting for this analysis is that AMICO does not make explicit use of colors, limiting the possibility of biasing the selection toward systems with a clear RS, which is crucial when we want to study its buildup process and the diversity in the galaxy population that clusters and groups show, for instance, at $z\gtrsim 1-1.5$. 
In this study, we also analyzed the first AMICO catalog generated on the COSMOS field  \citep{toni_amico-cosmos_2024} to explore the relationship between our results and the group X-ray emission. The sample covers a $\sim 3$ times larger area (the full COSMOS field) up to $z=2$ and includes mass estimates and X-ray properties of optically selected groups.

The paper is organized as follows. Section \ref{dataset} introduces the COSMOS group catalogs created with AMICO and the underlying galaxy catalogs. In Sect. \ref{ml}, we describe our machine learning (ML) method of classifying galaxies as quiescent or star-forming, including training, algorithm testing, and performance comparison. Section \ref{fredsec} analyzes the evolution of quiescent fractions in COSMOS groups and their dependence on group properties. Section \ref{redseq} characterizes the RS in our sample and compares it with evolutionary synthesis models and previous studies. Finally, Sect. \ref{conclusions} summarizes the main results and outlines future developments.
In our analysis, we assume a standard concordance flat $\Lambda$CDM cosmology with $\Omega_\mathrm{m}=0.3$, $\Omega_\Lambda =0.7$, and $h=H_0$ / (100 km/s/Mpc) $=0.7$. Magnitudes are expressed in the AB system and corrected for Galactic extinction.

\section{The catalog of galaxies and galaxy groups}\label{dataset}
The COSMOS field \citep{scoville_cosmic_2007} is one of the most data-rich extragalactic regions, and benefits from extensive multi-wavelength coverage from X-ray to radio \citep[e.g.,][]{hasinger_xmm-newton_2007, civano_chandra_2016, smolcic_vla-cosmos_2017}. Optical imaging encompasses coverage from several instruments, including the Canada-France-Hawaii Telescope \citep{sawicki_cfht_2019}, the Subaru Suprime-Cam \citep{taniguchi_subaru_2015}, the high-resolution Hubble Space Telescope ACS data \citep{koekemoer_cosmos_2007}, and the broadband data in the $g$, $r$, $i$, $z$, and $y$ filters from the Subaru Hyper Suprime-Cam \citep{aihara_third_2022}. Near-infrared coverage from the UltraVISTA survey \citep{mccracken_ultravista_2012, moneti_vizier_2023} further complements this extensive dataset.

Building on this legacy, the COSMOS-Web Survey \citep[PIs: Kartaltepe and Casey;][]{casey_cosmos-web_2023} is a JWST Cycle 1 program covering 0.54 deg$^2$ with four filters F115W, F150W, F277W, and F444W in NIRCam \citep{rieke_performance_2023,franco_cosmos-web_2025}, achieving 5$\sigma$ point-source depths of 27.5–28.2 mag. These filters offer a coverage of the near-infrared regime that is highly complementary to the UltraVISTA survey filters. An additional non-contiguous area is observed with the F770W MIRI filter \citep{wright_james_2022,harish_cosmos-web_2025}. 

The COSMOS-Web photometric galaxy catalog \citep[also known as COSMOS2025;][]{shuntov_cosmos2025_2025}, specifically targets the JWST-imaged region, building on previous COSMOS catalogs such as COSMOS2009 \citep{ilbert_cosmos_2009}, COSMOS2015 \citep{laigle_cosmos2015_2016}, and COSMOS2020 \citep{weaver_cosmos2020_2022}, and includes over 784,000 sources.The COSMOS2025 catalog sources were detected using a PSF-homogenized $\chi^2$ image combining all NIRCam bands, using \texttt{SourceXtractor++} \citep{bertin_sourcextractor_2022} to address PSF variations across datasets. This approach boasts a high spatial resolution as reflected in the deblending power of the source extraction.
In the COSMOS2025 catalog, redshifts (and all physical properties) are derived using the \texttt{LePhare} template-fitting code \citep{arnouts_measuring_2002, ilbert_accurate_2006} with an expanded set of templates based on \citet{bruzual_stellar_2003} models \citep{ilbert_evolution_2015}. When compared to high-confidence spectroscopic samples, the photometric redshifts achieve a precision of $\sim$ 0.01 for bright galaxies ($m_{F444W} < 23$) with $<2$\% catastrophic outliers, and maintain $<$ 0.03 precision with $\sim$10\% outliers even for $26<m_{F444W} < 28$. When divided between quiescent and star-forming, according to the NUVrJ criterion \citep{ilbert_mass_2013}, galaxy photo-$z$s achieve 0.008 and 0.013 precision, and 2.26\% and 1.95\% outliers, respectively  \citep[see Table 3 in][]{shuntov_cosmos2025_2025}.

We based our analysis on the new COSMOS-Web group and protocluster core catalog, presented in \citet{toni_cosmos-web_2025} and extending up to $z=3.7$ (the COSMOS-Web group catalog, hereinafter). The candidate groups were detected using the AMICO algorithm \citep[][]{bellagamba_optimal_2011, bellagamba_amico_2018,maturi_amico_2019}. 

AMICO is a linear optimal matched filter that detects clusters and groups in photometric galaxy catalogs using the galaxy position, photometric redshift, and an additional galaxy property (in the simplest case), such as the magnitude in a reference band. During the detection process, AMICO produces estimates of amplitude for each pixel of the analyzed three-dimensional space of sky coordinates and redshifts. The amplitude is the result of the convolution of the data with an optimal filter (for details, we refer the reader to \citealt{bellagamba_amico_2018} and \citealt{maturi_amico_2019}). Subsequently, the algorithm selects the candidates at the peaks in the amplitude map with the highest signal-to-noise ratio ($S/N$). After each selection, it assigns member galaxies to the cluster/group candidate by attributing a membership probability that depends on the galaxy properties, i.e., the magnitude and the redshift. Finally, AMICO uses this information to remove the signal of the detected object from the map, allowing easier detection of blended systems \citep[for details, see][]{bellagamba_amico_2018,maturi_amico_2019}. Each AMICO run returns a list of candidates with their respective lists of member galaxies, each with its association probability.

The COSMOS-Web group catalog was retrieved by applying AMICO to COSMOS2025. The group search was performed using the galaxy position, photo-$z$ distribution, and magnitude in a reference band, the NIRCam F150W band, extending to magnitude 27.3, making it the deepest AMICO application to date. The catalog includes 1678 group and protocluster core candidates in the COSMOS-Web field up to $z=3.7$, with $S/N>6.0$, and richness ($\lambda_\star$) cut at a minimum value of 2 and reaching up to $\sim 80$. More than 500 groups have their redshift confirmed by assigning spectroscopic counterparts \citep{toni_cosmos-web_2025}.

In this study, we also include an analysis of our first COSMOS catalog, AMICO-COSMOS \citep{toni_amico-cosmos_2024}, to explore the relation with group X-ray emission. Unlike COSMOS-Web, AMICO-COSMOS was built by applying the AMICO algorithm to the COSMOS2020 (and 2015) catalog (covering the full COSMOS field). This group search was tested with three different photometric bands. The catalog boasts the availability of associated X-ray properties, such as luminosity, flux, and mass, for over 600 candidate groups.

Given the sample richness estimates and the masses inferred by \citet{toni_amico-cosmos_2024} being in the range $M_{200} \approx 6 \times 10^{12} - 3 \times 10^{14} \, M_\odot$, in this work, we will deal mainly with galaxy groups rather than galaxy clusters. Only a few objects in the COSMOS field are expected to have masses larger than $10^{14}$ M$_\odot$ or more than 50 members\footnote{these are the commonly adopted thresholds used in the literature \citep[e.g.,][]{paul_understanding_2017,lovisari_scaling_2021} to separate groups and clusters} according to previous detections performed in the COSMOS field \citep[e.g.,][]{knobel_zcosmos_2012,gozaliasl_chandra_2019,toni_amico-cosmos_2024,toni_cosmos-web_2025}. Therefore, we will refer to our galaxy systems mainly as galaxy groups.

\section{Classification of galaxies with machine learning}\label{ml}
Since the AMICO galaxy group catalogs are based on input photometric galaxy catalogs in the COSMOS field, our goal is to classify COSMOS galaxies.
Several methods have been developed over the past years to select quiescent and star-forming galaxies \citep[see e.g.,][for an overview]{pearson_influence_2023}, spanning from a classic sharp cut in sSFR \citep{salim_dust_2018}, to color selections \citep[e.g.,][]{whitaker_newfirm_2011,ilbert_mass_2013} and the use of spectroscopic features \citep[e.g.,][]{baldwin_classification_1981,gallazzi_charting_2014}. 
In this work, in order to classify galaxies into quiescent (red) and star-forming (blue), we used a new approach that incorporates the information from several “classical” methods to train a ML model.
We chose to test two algorithms, one nonlinear and one linear: the \texttt{eXtremeGradientBoosting} classifier \citep[\texttt{XGB};][]{chen_xgboost_2016} and the \texttt{LinearDiscriminantAnalisys} classifier \citep[\texttt{LDA};][]{fisher1936lda}, both as implemented in the \texttt{scikit-learn} Python package \citep{pedregosa_scikit-learn_2011}.

\subsection{Machine learning algorithms}
The \texttt{XGB} algorithm belongs to the class of gradient boosted tree models \citep{friedman_greedy_2000}. In simple terms, this means it classifies data by building a series of decision trees, where each new tree learns from the mistakes made by the previous one. The correction of errors is based on a method called gradient descent optimization, which consists of iterative adjustments of the model parameters to “descend” toward the minimization of the prediction errors. The \texttt{XGB} algorithm is well suited for classification tasks on structured data, because it provides high predictive accuracy, built-in handling of missing values, and uses a technique called regularization -- a safeguard that prevents the model from fitting too closely to the training data (an issue known as overfitting). It also includes parameters to handle imbalanced class distributions effectively. Compared to linear classification methods, it can model complex, nonlinear relationships more effectively, despite being more computationally expensive and based on more complex modeling.

The \texttt{LDA} algorithm classifies data using linear combinations of features, assuming that targets in classes are normally distributed. This makes it a simple and efficient classification method, especially for small datasets for which these assumptions hold. However, it is less flexible than nonlinear methods such as \texttt{XGB}, less efficient with complex patterns, and does not include a default handling of imbalanced classes or missing values. For this application, we are dealing with highly imbalanced data, since photometric galaxy catalogs are expected to contain many more examples of blue star-forming galaxies rather than red quiescent galaxies. In the classification problem, these two are therefore referred to as the majority and minority classes, respectively.
Class imbalance can be mitigated through the generation of synthetic data. One of the most used methods for data generation is the Synthetic Minority Over-sampling Technique \citep[\texttt{SMOTE};][]{chawla_smote_2011}, which creates additional examples for the minority class via feature interpolation between existing data. This generally helps improve decision boundaries and reduces bias toward the majority class. We tested the classification with \texttt{LDA} both with and without the generation of synthetic data. The results of this test are shown in Table \ref{results_ml} and discussed in Sect. \ref{results_sect_ml}. The \texttt{LDA} method also requires the data to be scaled and normalized before training, which was done with the \texttt{StandardScaler} method from \texttt{scikit-learn}.

\begin{figure}
   \centering
   \includegraphics[width=9cm]{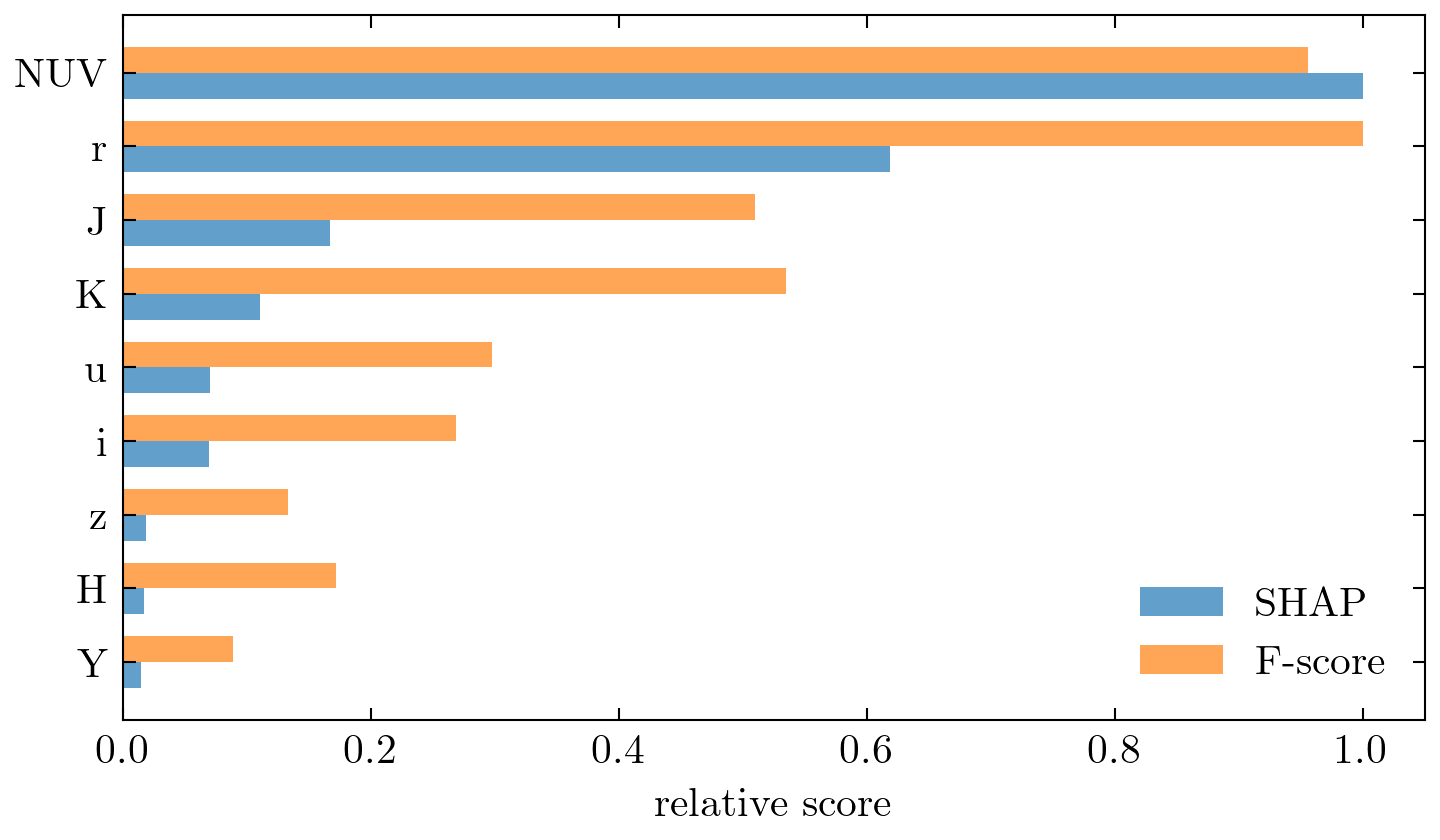}
      \caption{Feature importance for an initial \texttt{XGB} training, measured by the F score (relative split frequency; orange bars), highlights bands directly used to define ground-truth labels are the most frequent in decision splits. To better assess feature impact, we evaluate SHAP scores \citep{lundberg_unified_2017}, which quantify the relative feature contribution to predictions (blue bars). Both approaches confirm that rest-frame magnitudes in $r, \, NUV, \, K $, and $J$ provide the strongest constraining power.}
         \label{importance}
\end{figure}
\subsection{Training and testing dataset}\label{trainandtest}
We chose to train and test the ML models on a galaxy sample based on the COSMOS2015 galaxy catalog \citep{laigle_cosmos2015_2016}. Choosing this sample ensures a realistic representation of the features and properties of the galaxies that we want to classify. We chose not to train the algorithm on a subsample of COSMOS2025 since we aim to explore the full sample without compromising statistics and without introducing the risk of repeated data points in the training and target set, which may lead to overfitting. By choosing COSMOS2015 as the training set, we have no data-point overlap with the target dataset. Additional details about the training/testing set choice can be found in Appendix \ref{appendixA}.

The first step to create a training and testing set is to clean the parent catalog to reduce the possibility of training the models on misclassified or spurious objects. For this reason, we performed a conservative cleaning applying the following cuts:
\begin{itemize}
    \item we removed stars (\texttt{type=1}) and  unclassified objects due to SED fitting failure;
    \item we selected only data with high-quality photometry according to the \texttt{SExtractor} flagging system in optical broad bands and UltraVISTA bands: $\texttt{[BAND]\_FLAGS}<4$;
    \item we kept only safe and unmasked objects, namely those with \texttt{FLAG\_PETER in (0,4,6)} (for details on this selection, see Sect. 2 of \citeauthor{toni_amico-cosmos_2024} \citeyear{toni_amico-cosmos_2024}) and additionally removed any objects falling inside the AMICO-COSMOS visibility mask, described in Sect. 4.3 of \citet{toni_amico-cosmos_2024};
    \item we removed galaxies located within 0.05 degrees of pairs of a bright star halo (or a halo-border pair), resulting in a more conservative mask where the field is more fragmented;
    \item finally, we cut the dataset at $H<25$ (corresponding to the mode of the magnitude distribution) to discard faint sources with large magnitude errors and homogenize completeness.
\end{itemize}
This cleaning yielded a sample of approximately 300,000 galaxies in our training/testing set, across the range $0<z<6$.
In order to establish the “ground truth” in the classification of galaxies for the training/testing set, we used the classification label yielded by several classical methods. We proceeded as follows. First, we selected four classical methods widely used in the literature to discriminate between the two galaxy classes:
\begin{itemize}
    \item NUVrJ \citep{ilbert_mass_2013}: a rest-frame color-color cut at $NUV -r = 3(r - J) + 1$ and $NUV - r = 3.1$;
    \item sSFR threshold \citep{salim_dust_2018}: a fixed cut in sSFR at $10^{-10.5} \, yr^{-1}$, as adopted for instance by \citet{euclid_collaboration_humphrey_euclid_2023} and \citet{bisigello_euclid_2020};
    \item Sa-color evolution \citep{andreon_build-up_2006, radovich_amico_2020}: a cut given by the evolution in $z$ of the observed color of a modeled Sa-type galaxy. Colors bracket the 4000 $\AA$ break;
    \item NUVrK \citep{davidzon_vimos_2016}: a rest-frame color-color cut at $NUV - r = 1.37(r - K) + 2.6$ and $NUV - r = 3.15$ and $r - K = 1.3$.
\end{itemize}
We classified galaxies into three categories according to these four methods: the galaxy is flagged as quiescent if it satisfies most of the four criteria (7\% of the sample). The same procedure is followed to classify star-forming galaxies (92\% of the sample). If the classification is ambiguous (e.g., 2 vs. 2), the galaxy is stored in the third class. 
This third class allows us to identify galaxies with the following characteristics: those with uncertain classifications, because of missing UV photometry and/or ambiguous SED fitting, as well as those in a physical transition phase between the RS and the blue cloud, a regime commonly referred to as the green valley. This category of objects represents $\sim 1\%$ of the full sample. Given the small statistics, it could not be introduced as a third class in the learning process and the classification. Therefore, we decided to remove these objects from the training and testing sets. This reduces ambiguity in the classification and results in an improvement of the performances we describe in Sect. \ref{results_sect_ml} of about $2-5\%$ in all evaluation metrics for the minority class.
The final selected sample was then randomly divided into training (2/3) and testing set (1/3) using the \texttt{scikit-learn} \texttt{train\_test\_split} function. Additional information about the classical method thresholds, their properties, and the distribution of labeled galaxies in redshift and mass can be found in Appendix \ref{appendixA}.

\subsection{Feature engineering }
In ML, features are the measurable properties available in the data that we select to make predictions on the objects we want to target. Feature engineering is the ML pre-processing step of selecting, evaluating, and adjusting features, including the handling of missing data, so that the ML model can better capture patterns in the data to make predictions.
First of all, we selected all rest-frame magnitudes available in both COSMOS2015 and COSMOS2020. These are the magnitudes in the following bands: $NUV, \, u, \, r, \, i, \, z, \, Y, \, J, \, H$ and $ K $.
To get an estimate of the impact of features on the decision-making process, we performed an initial training with the \texttt{XGB} algorithm, which easily allows the evaluation of feature importance. The results are shown in Fig. \ref{importance}. The importance is measured in terms of the F score (also known as weight, displayed by the orange bars), which is the number of times a feature appears in the decision tree splitting. The $r$, $NUV$, $J$, and $K$ bands are the most frequent features used in the decision-making process. This result is expected, as these bands are directly used in assigning the ground-truth labels, as described in Sect. \ref{trainandtest}. To gain deeper insight into feature impact, we also evaluated the SHAP score \citep{lundberg_unified_2017}, which indicates how much each feature contributes to the model’s predictions (blue bars). The rest-frame magnitudes in $r$, $NUV$, $J$, and $K$ are confirmed as the features with the greatest constraining power in the model predictions. These four magnitude bands are also the only rest-frame magnitudes available in the current release of the COSMOS2025 galaxy catalog at the moment of writing. Since the selection of features is not trivial, and often a few features with high importance simplify the model and lead to better performance than a large number of features, we chose to test and compare results using all magnitudes and with only the best-ranked four. In Sect. \ref{results_sect_ml}, we describe the differences in performance depending on the set of chosen features in more detail.
\begin{figure}
   \centering
   \includegraphics[width=9cm]{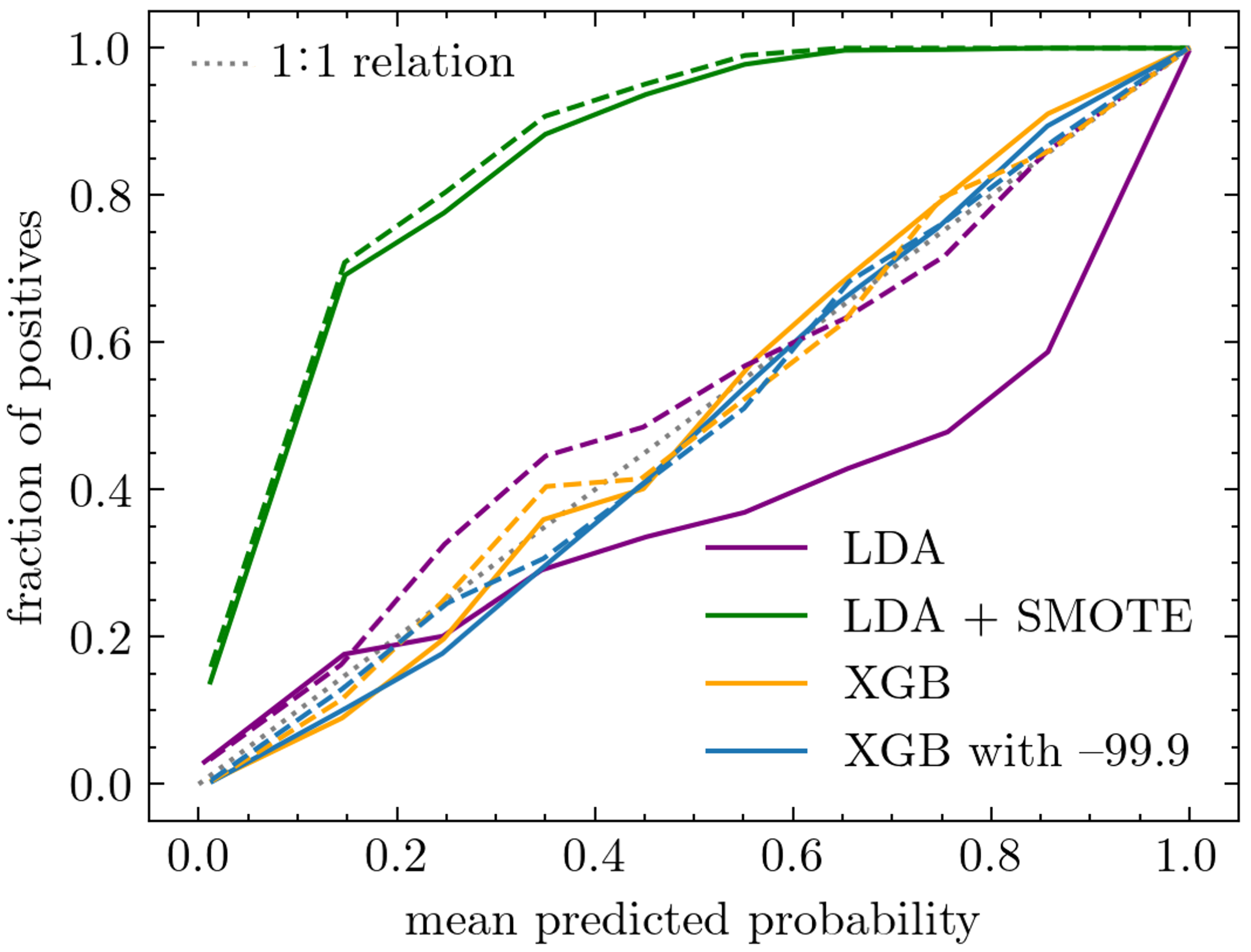}
      \caption{Reliability curves showing the bias between predicted probability and fraction of actual positive labels (star-forming galaxies) using all rest-frame magnitude bands (solid lines) and the four most impacting in the decision (dashed line). Different colors represent different methods and configurations, as in the legend, and the dotted gray line represents the ideal one-to-one relation.}
         \label{prob_calibration}
\end{figure}
\begin{figure*}
   \centering
   \includegraphics[width=18cm]{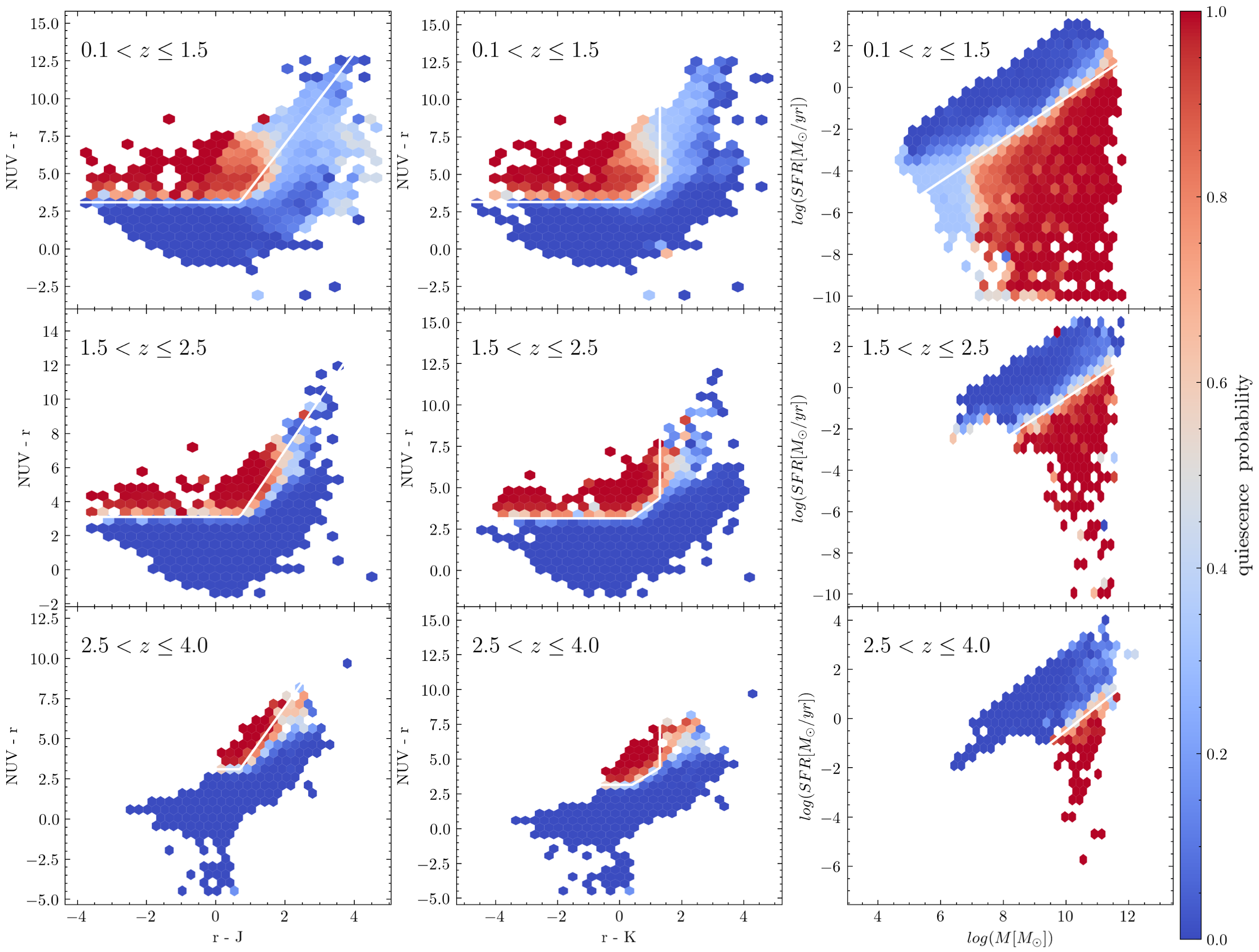}
      \caption{Results of our probabilistic classification of COSMOS-Web galaxies with \texttt{XGB} with imputation, as in the NUVrJ (left column), NUVrK (middle column), and SFR--$M$ (right column) planes. Each hexagon is color-coded by the mean quiescent probability (scale bar on the right), with rows corresponding to different redshift bins. As expected from our training, the method consistently reproduces classical cuts (white lines), while also providing additional insight near the classification boundaries. Classification is ambiguous for some low-mass galaxies at low redshift (light blue region in the top panels).}
         \label{classif}
\end{figure*}

The \texttt{XGB} algorithm is sparsity-aware, meaning it can naturally handle missing data. This can be particularly useful in regions with reduced photometric coverage, as it allows us to include galaxies that are missing some rest-frame magnitudes rather than discarding them completely. Missing values are imputed with an arbitrary constant (--99.9), chosen to be clearly distinguishable from real values. Because \texttt{XGB} does not create artifacts from such artificial flags, this strategy is effective; indeed, \citet{euclid_collaboration_humphrey_euclid_2023} showed that using a flagging number, like --99.9, outperforms other imputation methods such as the mean, median, or minimum. This approach enables the algorithm to learn from both fully and partially characterized galaxies. We tested the \texttt{XGB} algorithm with and without imputation of a constant value to recover missing data.

\subsection{Classification performance and results}\label{results_sect_ml}
To evaluate the algorithms and configurations, we relied on the standard metric used to evaluate classification performance against the testing set: precision, recall, and F1 score. Precision ($P$) is the fraction of correctly identified objects among all objects assigned to a class, analogous to purity in astronomical samples. Recall ($R$) is the fraction of true objects correctly identified, analogous to completeness. The F1 score combines both precision and recall via their harmonic mean (i.e., $2PR/P+R$). These three quantities range between 0 and 1.

\begin{table}[h]
      \caption[]{Performance of the classification methods in terms of precision, recall, and F1 score. }
         \label{results_ml}
     $$ 
         \begin{tabular}{c  c c c c}
            \hline
            \noalign{\smallskip}
            Method & Precision & Recall & F1 score \\
            & \small{(red/blue)}& \small{(red/blue)} & \small{(red/blue)} \\
            NUV,$u,r,i,z$,Y,J,H,K & & & & \\
            \hline
            \noalign{\smallskip}
            \texttt{LDA} & 0.94/0.99 & 0.84/1.00 & 0.89/0.99 \\
            \noalign{\smallskip}
            \texttt{LDA + SMOTE} & 0.69/1.00 & 1.00/0.97 & 0.81/0.98 \\
            \noalign{\smallskip}
            \texttt{XGB} & 0.96/1.00 & 0.95/1.00 & 0.95/1.00 \\
            \noalign{\smallskip}
            \texttt{XGB} with --99.9 & 0.96/1.00 & 0.93/1.00 & 0.94/1.00 \\
            \noalign{\smallskip}
            \noalign{\smallskip}
            NUV,$r$,J,K & & & & \\
            \hline
            \noalign{\smallskip}
            \texttt{LDA} & 0.90/0.99 & 0.90/0.99 & 0.90/0.99 \\
            \noalign{\smallskip}
            \texttt{LDA + SMOTE} & 0.66/1.00 & 1.00/0.96 & 0.80/0.98 \\
            \noalign{\smallskip}
            \texttt{XGB} & 0.96/1.00 & 0.95/1.00 & 0.95/1.00 \\
            \noalign{\smallskip}
            \texttt{XGB} with --99.9 & 0.95/0.99 & 0.91/1.00 & 0.93/1.00 \\
            \hline
         \end{tabular}
     $$ 
     \tablefoot{The upper block refers to the training with all available rest-frame magnitude bands, the bottom block to the one using only the four with the highest decision impact. 
     Testing set sample size and red/blue ratio: 94236 galaxies with 7.1\% red/blue ratio; 98659 with 7.2\% red/blue ratio, when missing values are imputed.}
   \end{table}

The performances evaluated against the testing set are summarized in Table \ref{results_ml}, in terms of the metrics just introduced, both for the minority (quiescent galaxies, red) and for the majority class (star-forming galaxies, blue).
The different methods yield comparable results for the classification of star-forming galaxies, whereas for the identification of quiescent galaxies, the \texttt{XGB} algorithm tends to perform better overall. 
Interestingly, applying \texttt{SMOTE} to help \texttt{LDA} with class imbalance increases recall for the minority class but significantly reduces precision. This may be due to an overlap between the two class distributions in the feature space, happening when classes share some similar properties and boundaries are not clearly separable, which is expected to cause overconfidence toward the minority class when synthetic data are injected \citep[e.g.,][]{chawla_smote_2011,lu_overlapping_2024}.

However, in this application, we are concerned not only with the quality of the binary classification of our target sample, but particularly with the reliability of the probability assigned to each object of belonging to a given class, since these probabilities are used to estimate quiescent fractions and identify the RS.
Therefore, in addition to recall, precision, and F1 score, we analyzed how well the probabilities are estimated. Figure \ref{prob_calibration} shows the so-called reliability curve (also known as calibration curve), which plots the fraction of actual positives (in this case, “true” star-forming galaxies) against the mean predicted probability for that class, for all four methods, in bins of predicted probability. This diagnostic illustrates how closely the predicted probabilities match the true outcomes. A reliable, unbiased model produces a curve that follows the one-to-one diagonal in the plot: for example, if the mean predicted probability in a given bin is 0.7, then roughly 70\% of the objects in that bin should belong to that class. 

One case in which good classification performance does not imply good calibration is the model trained using data augmentation to balance the classes. Addressing the imbalance between star-forming and quiescent galaxies can lead the model to underestimate the probabilities of the majority class (positives), producing a reliability curve well above the one-to-one line (see the green curves in Fig. \ref{prob_calibration}). The reliability curves also reveal the sensitivity of \texttt{LDA} to the number of selected features. For simple linear models, using a smaller set of strong features can yield better results than relying on a more complex set of galaxy properties.

By implementing the imputation of the missing values instead of complete discard, we observed a slight decrease in the performance for minority class recall, but a more reliable probability estimate. For this application, given the importance of an unbiased probability and the stability of the algorithm with respect to the number of features used, we decided to opt for an \texttt{XGB}-based classification with the imputation of a constant value for missing data. In Fig. \ref{prob_calibration}, this configuration is represented by the dashed blue line, which is the most stable and closest to the one-to-one relation.

The results of the application of our chosen model for the probabilistic classification, to the COSMOS2025 dataset used for the COSMOS-Web group search, are shown in Fig. \ref{classif}: each hexagon is color-coded according to the mean probability to be quiescent (bar on the right) inside that hexagonal bin; each column shows a different diagram used in classical methods, while rows refer to different redshift intervals. As expected from the labeling of our training set, our method is consistently mapping classical empirical cuts (displayed as white lines), while providing additional insight, particularly in regions near the classification boundaries. The algorithm struggles in the classification of galaxies with masses below $10^7 M_\odot$ in the first redshift bin, populating the light blue region in the top panels. These galaxies mostly have $z<0.2$, and are expected to have a small impact on our analysis, as we are particularly interested in evolution at earlier cosmic times.

In a recent parallel study, \citet{asadi_machine_2025} classified COSMOS2020 galaxies \citep{weaver_cosmos2020_2022} from the \texttt{Farmer} catalog release \citep{weaver23} using a gradient boosted tree algorithm trained on a mock galaxy catalog based on semi-analytical models and employing 28 colors. Their approach differs from ours in several key aspects: it relies on simulated rather than observed data for training, uses apparent colors instead of rest-frame magnitudes, and labels the training set solely through an sSFR threshold. Since their trained model is publicly available, we applied it to the new COSMOS-Web data. The outcomes align with their findings for COSMOS2020: specifically, their model returns a higher fraction of quiescent galaxies at all redshifts compared to classical thresholds such as NUVrJ. Because our method is based on the consensus among multiple classical thresholds (see Sect. \ref{trainandtest} and Appendix \ref{appendixA}), the model of \citealt{asadi_machine_2025} also predicts a consistently higher quiescent fraction relative to the approach presented in this work.

\section{Quiescent galaxy fractions}\label{fredsec}

We studied the fraction of quiescent (red) galaxies in groups and its dependencies on group richness and redshift. The quiescent fractions were computed with two different methods: pure membership (with testing of secondary association subtraction) and cylinder background subtraction.

Following common practice to reduce biases due to color uncertainties at the faint end and fully sample typical ranges occupied by RS galaxies, we introduced a redshift-dependent luminosity limit to define the quiescent fraction. This limit is often defined at 2 or 3 magnitudes fainter than an evolving characteristic magnitude or the magnitude of a reference galaxy, such as the brightest one \citep[e.g.,][]{stott_evolution_2009,cerulo_accelerated_2016,radovich_amico_2020}. The depth of COSMOS-Web allows us to push this limit deeper than in previous studies and to adopt a threshold at $m_\star(z_j)+4$, where $z_j$ is the redshift of the group. In other words, for each group, we analyzed all galaxies up to 4 magnitudes fainter than the magnitude of a typical elliptical galaxy at the knee of the group luminosity function, $m_\star(z_j)$, in our reference band (NIRCam F150W), which was estimated by evolving a massive passively evolving elliptical galaxy \citep[see][for details]{toni_cosmos-web_2025} with evolutionary synthesis models \citep{kotulla_galev_2009}. This limit allows for a deep sampling of sub-$m_\star$ galaxies (galaxies fainter than the luminosity function knee), which include the strongest signatures of RS buildup \citep[e.g.,][]{stott_increase_2007,stott_evolution_2009}. It should be noted that the $m_\star +4$ limit is not a sharp cut for the galaxy sample, because it depends on the redshift of the group the galaxies are associated with, so galaxies are limited in general by the survey's completeness limit \citep[see][]{shuntov_cosmos2025_2025, toni_cosmos-web_2025}. However, at $z \sim 2.5$, the evolving $m_\star(z_j)$ reaches the sample magnitude limit. Beyond this redshift, the $m_\star+4$ cut is no longer significant, because the faint-end is determined by the survey’s limit. This implies that some of the faintest galaxies in our last two redshift bins -- in particular, low-luminosity star-forming galaxies -- might be undetected because beyond the survey limit, which could lead to a slight overestimation of the quiescent fraction in those bins with respect to lower redshifts. However, as discussed later, this effect appears not to be significant in our sample, indicating that this selection yields a consistent population across cosmic time while fully leveraging the depth and statistical power of COSMOS-Web. In this section, we briefly introduce the three approaches to compute the quiescent fraction and present the results for this group sample.
\subsection{Pure membership and secondary association subtraction}\label{basic_fred}
Our sample includes the probability of each galaxy belonging to a given group, as provided by the AMICO algorithm. Moreover, we have the probability of each galaxy being quiescent as we estimated in Sect. \ref{ml}. The simplest way to compute quiescent fractions is by summing the probabilities to be quiescent, conditional on the probability of belonging to a group, and dividing it by the total, regardless of the class attributed in Sect. \ref{ml}. Given that $P_{i,j}$ is the AMICO probability of the $i$-th galaxy of belonging to the $j$-th group \citep[see][for details]{maturi_amico_2019} and that $P_{red,i}$ and $P_{blue,i}$ are the (complementary) probabilities of the galaxy to be quiescent or star-forming, respectively, the quiescent fraction of the $j$-th group, $f_{q, j}$, can be estimated as
\begin{equation}\label{fred}
    f_{q, j} = \frac{\sum P_{red,i} P_{i,j}}{\sum P_{red,i} P_{i,j}+\sum P_{blue,i} P_{i,j}} = \frac{\sum P_{red,i} P_{i,j}}{\sum  P_{i,j}} \, .
\end{equation}

This computation relies on AMICO membership probabilities, which are model-dependent, computed during the detection process and based on the chosen model of that specific group search, which is built on an NFW profile \citep{navarro_universal_1997} and a Schechter luminosity function \citep{schechter_analytic_1976}, for typical low-mass clusters or large groups \citep{toni_amico-cosmos_2024,toni_cosmos-web_2025}.
The AMICO association probability assignment also takes into account the possibility for a galaxy to be associated with multiple group detections. Thus, we introduced a small correction in the computation, which consists of considering only the main (highest probability) association for each galaxy and subtracting from it the probabilities to be associated with other detections (secondary associations). The probability, $P_{i,j}$, in Eq. \ref{fred} is therefore substituted by
\begin{equation}
    P_{i,j - new} = P_{i,j - main} - \sum_{N_{assoc}-1} P_{i,j - secondary} \, .
\end{equation}
However, this correction turned out to be minimal, and the results with and without correction are statistically compatible.

\begin{figure}
   \centering
   \includegraphics[width=9cm]{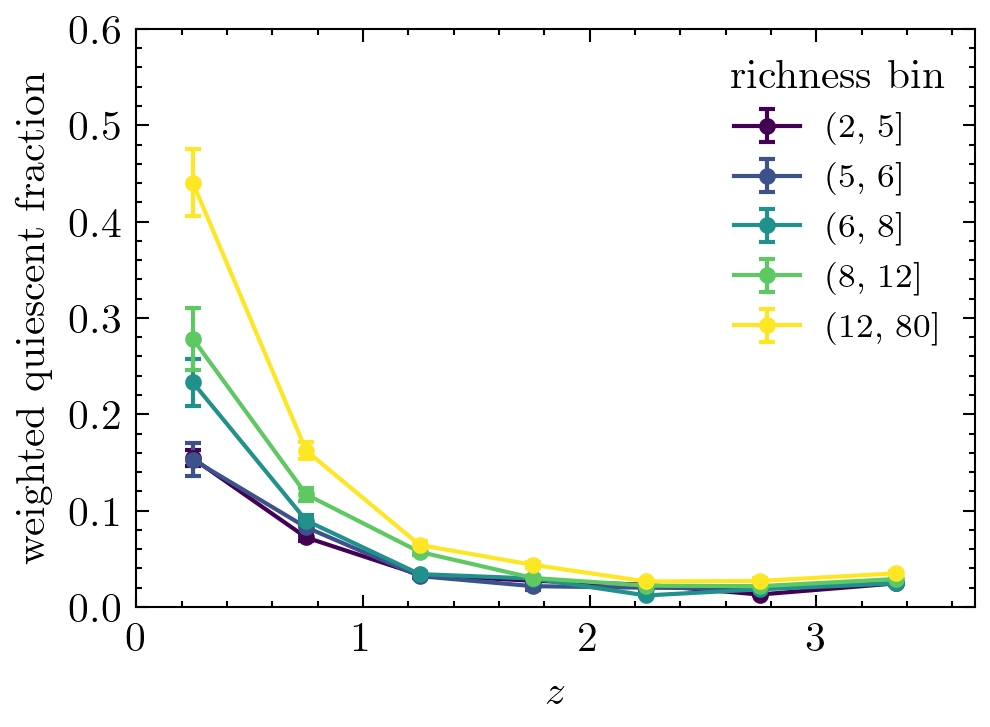}
      \caption{Purity-weighted quiescent fraction vs. redshift in different richness ($\lambda_\star$) bins, as in the legend. The richer the group, the faster its quiescent population buildup.}
         \label{basic_fred_fig}
\end{figure}

The results are reported in Fig. \ref{basic_fred_fig}, where we show the weighted quiescent fraction, that is, the quiescent fraction expressed by Eq. \ref{fred}, weighted by the purity of the sample in the corresponding redshift and richness bin. Purity was estimated against realistic data-driven mocks we have generated with SinFoniA \citep{maturi_amico_2019}, which exploits a Monte Carlo approach. All details about the mock generation and the purity and completeness estimation can be found in \citet{toni_cosmos-web_2025}. For comparison, we included the plot in Fig. \ref{basic_fred_fig} without purity weightning in Appendix \ref{appendixB}. 

Figure \ref{basic_fred_fig} shows that the population of quiescent galaxies seems to start assembling consistently shortly before $z=2$, with earlier and accelerated buildup for the richest systems. The slight increase in the quiescent fraction observed at $z \gtrsim 3$ may be influenced by the survey’s magnitude limit, as previously discussed. However, this rise is minimal and also consistent with the expected contribution from a few rare, early RS galaxies that may already be emerging at these high redshifts (see Sect. \ref{obsrs}). For this and the subsequent analyses, we chose richness bin sizes to homogenize the number counts whenever possible, and we estimated uncertainties using bootstrap resampling with 1000 realizations.

\subsection{Cylinder background subtraction}\label{cylsection}
The AMICO algorithm produces model-dependent membership probabilities for the group member galaxies. In some cases, when quantities used to model the distribution of galaxies in the template are studied, this might lead to biased considerations due to the underlying model chosen for the detection. This is the case for radial profiles and luminosity functions \citep[see e.g.,][]{puddu_amico_2021}. In \cite{toni_cosmos-web_2025}, we proposed a new method based on a cylinder-based background subtraction technique to retrieve model-independent results for individual and stacked analyses. For the study of the quiescent fraction, we tested the cylinder method, performing background subtraction and estimating the quiescent fraction in a model-independent way. The implementation proceeds as follows:
\begin{enumerate}
    \item The first step is the definition of cylindrical volumes around each group. We chose to make the cylinder extend $\pm$0.01(1+$z_{j}$) from the group center in the redshift direction, where $z_{j}$ is the group redshift. On the sky-plane, we chose a variable radius given by the root mean square (RMS) of the group-centric galaxy distribution. The choice of the cylinder parameters is described in detail in Appendix \ref{appendixB}. We call the volumes inside the cylinders “group regions.”
    \item Then, we must consider that not all volume is filled with galaxies, as areas can be masked due to the presence of bright stars and/or image artifacts and patterns. We estimated effective areas on the entire field (i.e., subtracting masked areas), using the same mask adopted in the detection process in each redshift slice (using the AMICO detection redshift resolution, $\Delta z =0.01$). We accounted for masked areas both inside and outside the cylinders, namely for “group regions” and “field regions,” respectively.
    \item We estimated quiescent and star-forming galaxy counts as the sum of the probability to be quiescent ($P_{\text{red}}$) or star-forming ($P_{\text{blue}}$), weighted by the value of the galaxy redshift probability distribution at the group redshift. Thus, the red and blue counts are expressed by
    \begin{equation}
       N_{red} = \sum_{i} P_{\text{red}, i} \, p_i(z_j) \,, \hspace{0.2cm}  \text{and} \hspace{0.2cm} 
       N_{blue} = \sum_{i} P_{\text{blue}, i} \, p_i(z_j) \, ,
    \end{equation}
    where the $p_i(z_j)$ is the redshift probability of the $i$-th galaxy at the redshift of the host group detection ($z_j$).
    \item We did this computation for galaxies in both group and field regions and normalized counts by the volumes ($V_{(1)}$ and $V_{(0)}$, respectively), in all redshift slices crossed by the cylinder;
    \item The final galaxy density within a group is the density in that group region minus the density in the field region at that redshift. Given that the subscripts (1) and (0) refer to galaxies and volumes in the group and field regions, respectively, the final density of quiescent (red) galaxies (equivalent for the star-forming, or blue) in the $j$-th group is given by
    \begin{equation}\label{bgsred}
        \rho_{\text{red},j} = \sum_{(1)} P_{\text{red}, i} \, p_i(z_j) \, \, V_{(1)}^{-1} 
  - \sum_{(0)} P_{\text{red}, i} \, p_i(z_j) \, \, V_{(0)}^{-1} \, 
    .\end{equation}
    \item Finally, the quiescent galaxy fraction is defined as the background-subtracted density of quiescent galaxies (Eq. \ref{bgsred}) over the total density. Galaxies and respective volumes located in the cylinder redshift slices, outside the cylinder volume, and along the line of sight of it, are rejected and not included in the calculation to avoid contamination. Thus, the final expression for the cylinder quiescent galaxy fraction, $f'_{\text{q}, j}$, is given by
\begin{align}\label{final_ratio}
f'_{\text{q}, j} 
&= \frac{
\rho_{\text{red},j}}{\rho_{\text{red},j} + \rho_{\text{blue},j}} \\
&= \frac{\rho_{\text{red},j}}{
  \sum_{(1)} p_i(z_j) \, V_{(1)}^{-1}
  - \sum_{(0)} p_i(z_j) \, V_{(0)}^{-1}
} \, .
\end{align}
\end{enumerate}

\begin{figure}
   \centering
   \includegraphics[width=9cm]{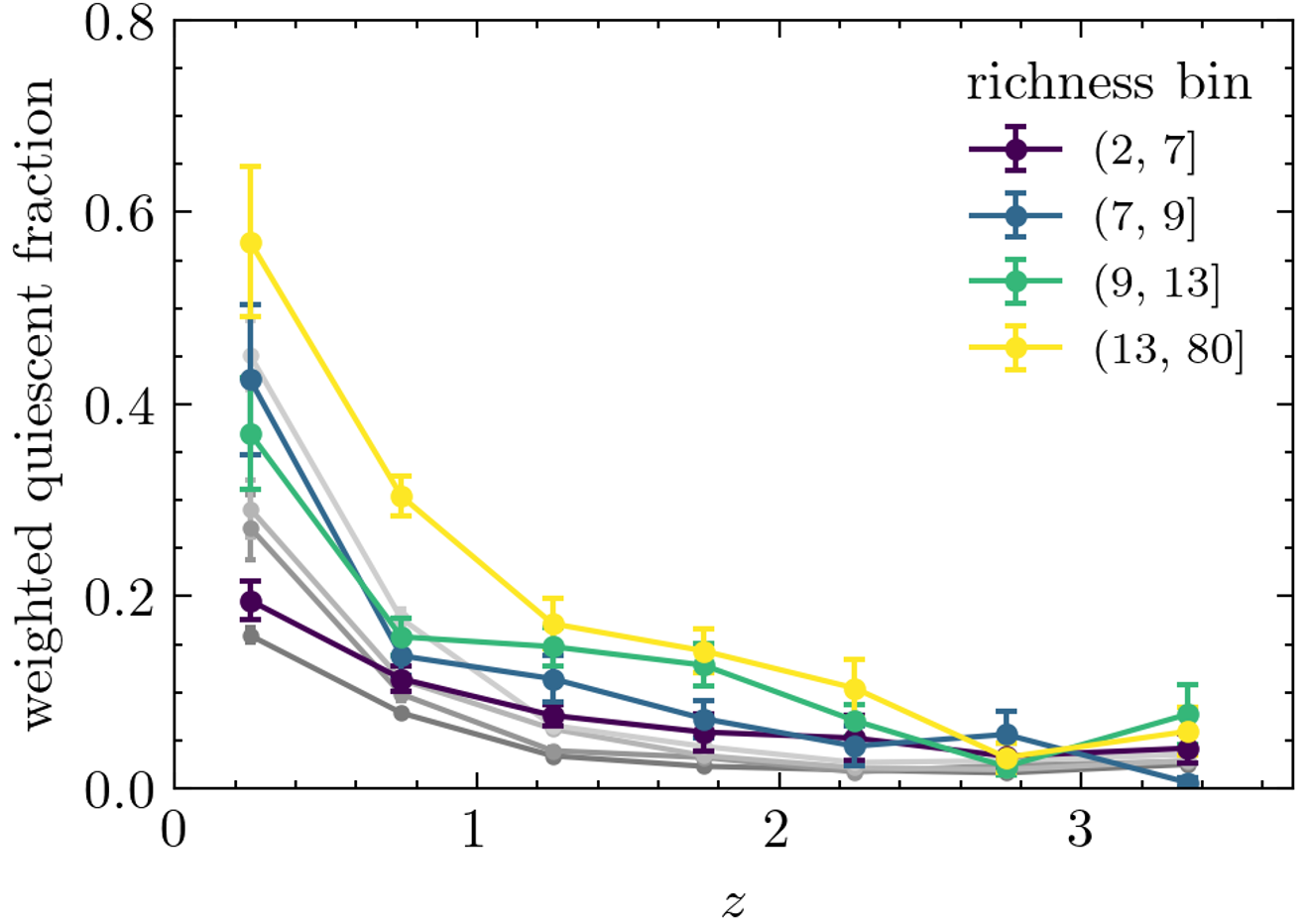}
      \caption{Purity-weighted quiescent fraction as estimated with the cylinder background subtraction, as a function of redshift. Different colors show different richness bins, as in the legend. The grayscale trend lines refer to the fractions obtained with the pure membership method in corresponding richness bins (Fig. \ref{basic_fred_fig}).}
         \label{fred_cyl}
\end{figure}

Cylinder-based quiescent fractions were computed only for groups with a fraction of masked area not larger than 40\% to reject extremely masked detections. We also removed groups in the extreme redshift bins to avoid border effects when considering the cylinders. Additional discussion about this computation and the method we used to deal with possible negative densities can be found in Appendix \ref{appendixB}.
To retrieve these results, as well as those using the membership probability, we weighted the quiescent fraction obtained as just described by the value of the purity in the corresponding bins of redshift and richness. Results without purity-weighting are reported in Appendix \ref{appendixB}. 
Results for the cylinder-extracted quiescent fractions are shown in Fig. \ref{fred_cyl} for four different bins of richness. For comparison, we report in grayscale the results for the pure-membership quiescent fractions in the corresponding richness bins (Fig. \ref{basic_fred_fig}). Apart from having larger uncertainties due also to lower statistics and to the scatter given by the background, the trend with redshift of the cylinder-based fractions is in line with a rapid buildup of the quiescent population starting at $z\sim2$, and $z \sim2.5$ for the richest groups, as seen when using pure memberships. However, when present, the quiescent population seems to be more dominant in rich groups when estimated through the cylinder-based method, which may be interpreted as a consequence of an effective field-galaxy contamination removal or an effect of background fluctuation, since the cylinder method retrieves fractions in comparison to the field population at that redshift. The discrepancy is less prominent in the low-richness bins. In conclusion, the model-dependent pure-membership method and the nearly model-independent\footnote{It should be noted that, although AMICO does not explicitly use color selections like RS-based detection algorithms, biases may still arise from the intrinsic properties of the data. In particular, quiescent galaxies tend to be massive and systematically brighter than star-forming ones. Even after removing model (and therefore magnitude) dependence via the cylinder method, the analysis may remain affected by the generally higher photo-$z$ accuracy of quiescent galaxies compared to star-forming ones.} cylinder method, despite their different assumptions and computation procedures, produce strikingly consistent trends over nearly 12 Gyr of cosmic time.

\subsection{X-ray luminosity of groups and environmental effects}

In recent years, growing attention has been dedicated to the observed differences and expected biases between X-ray- and optically selected samples of galaxy clusters and groups. It is now increasingly recognized that X-ray–bright and X-ray–faint (or X-ray-undetected) systems often display distinct physical properties, indicating that different selection methods can preferentially identify different underlying populations. These discrepancies trace physical signatures of processes such as AGN feedback and halo assembly history, commonly referred to as halo assembly bias, and their impact becomes especially significant when transitioning to the galaxy group regime \citep[e.g.,][]{lovisari_scaling_2021,eckert_feedback_2021,popesso_x-ray_2024,marini_impact_2025}.

\begin{figure}
   \centering
   \includegraphics[width=9cm]{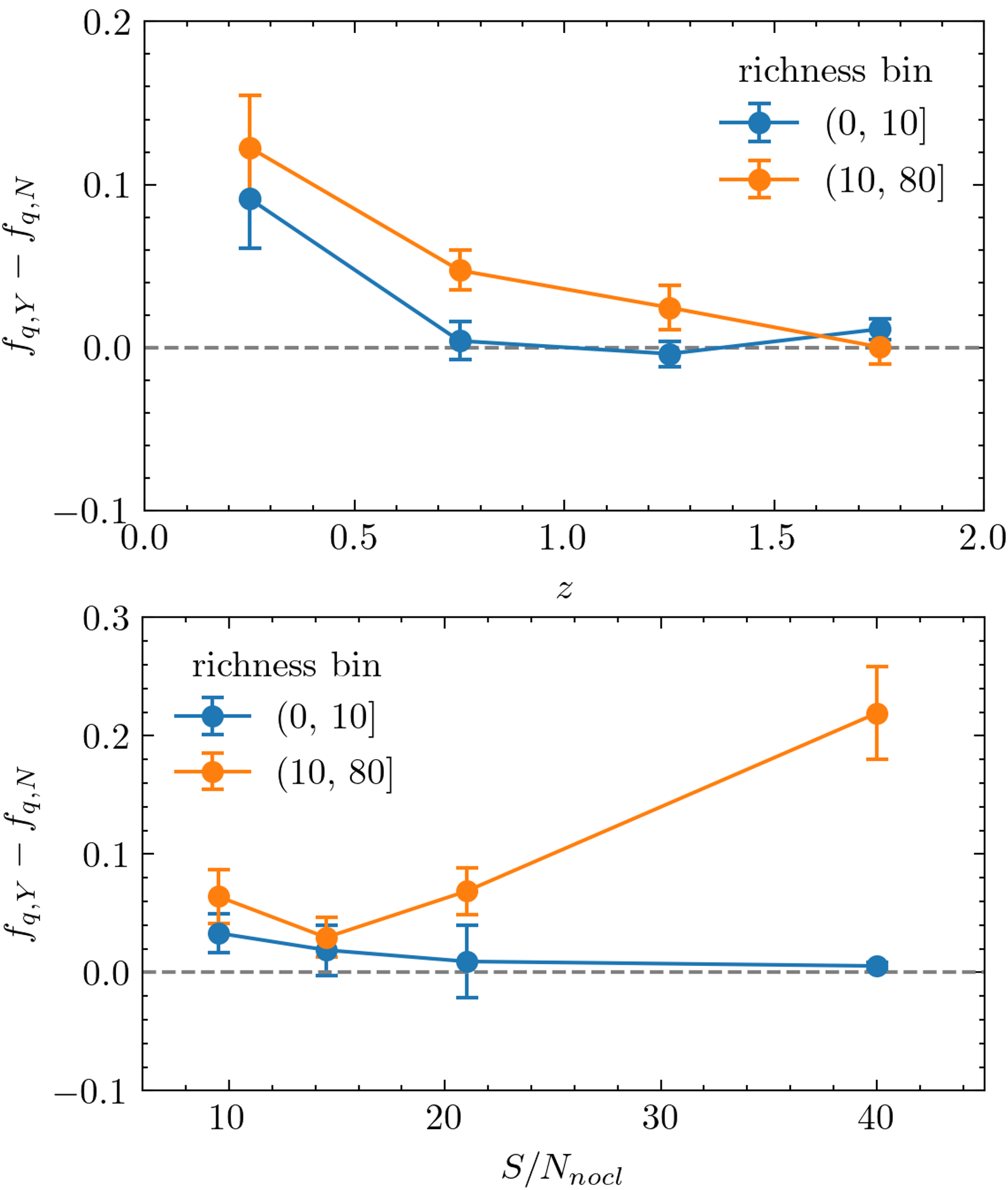}
      \caption{Quiescent fraction difference between X-ray bright ($Y$) and X-ray faint ($N$) groups as a function of redshift (top panel) and AMICO $S/N_{nocl}$ (bottom panel), in two bins of richness as in the legend.}
         \label{frac_x}
\end{figure}

In this Section, we consider the AMICO-COSMOS group sample \citep{toni_amico-cosmos_2024} comprising 622 candidate groups with X-ray flux above the significance level. Among these, 222 detections have a robust estimate of X-ray properties that was used to estimate group mass and to calibrate mass-proxy scaling relation in \citet{toni_amico-cosmos_2024}. In this sample, around half of the detections have no significant X-ray emission, and this is fairly independent of $S/N$ or redshift cuts. 
 We leverage the availability of X-ray flux and X-ray flux significance (i.e., $\sigma_X$) in this sample to study the quiescent fraction and its evolution for the population of X-ray bright groups and those for which a significant X-ray emission was not detected. We calculated the quiescent fraction with the pure membership method, for simplicity, dividing it into the population of robust X-ray emitters ($\sigma_X > 1$ and good quality flag from \citeauthor{toni_amico-cosmos_2024} \citeyear{toni_amico-cosmos_2024}; 164 groups) and the population with X-ray flux under the significance limit ($\sigma_X \leq 1$; 359 groups). The difference between quiescent fraction for the two populations is shown in Fig. \ref{frac_x} as a function of the redshift (top panel) and AMICO signal-to-noise ratio, $S/N_{nocl}$ (bottom panel), for two different richness bins (blue and orange lines). The result is that groups with clear X-ray emission tend to have a quiescent fraction consistently higher than or at most equal to that of groups without significant X-ray emission. In particular, rich groups detected with high $S/N$ and groups at $z \lesssim 0.5$ seem to have significantly more dominant population of quiescent galaxies when they are bright in the X-rays.

In a recent study, \citet{popesso_x-ray_2024} compare the GAMA and eROSITA samples by analyzing the properties of X-ray-undetected clusters and groups. Among the considered properties, they study the position of the systems with respect to the large-scale structure (LSS), finding that $\sim$90\% of low-emission clusters and groups, often undetected in the X-rays, are located in filaments (and sheets) rather than in the nodes of the cosmic web. Interestingly, this might be correlating with the properties of the hosted galaxy population, like the pre-accretion processing and therefore the star formation activity. Groups residing in nodes are more likely to accrete galaxies pre-processed in filaments, and therefore are more quenched on average. On the other hand, groups hosted by filaments accrete galaxies from the field that are therefore less processed.
To check whether this is the case for our sample as well, we compared the position of our groups with that of galaxies that have been found to be located in filaments or in clusters based on local density in LSS maps. For this, we used data from the COSMOS LSS study by \citet{darvish_cosmic_2014}\footnote{The density field was quantified through a scale-independent Multi-scale Morphology Filter (MMF) algorithm, on which the extraction of LSS components (field, filaments and clusters/nodes) was based (see \citealt{darvish_cosmic_2014}, Sect. 3 for details).}, which covers the redshift interval $0.1 \leq z \leq 1.4$ and the more recent data based on the COSMOS2020 galaxy catalog, which extends from $z=0.5$ to beyond the higher end of our group sample (Taamoli et al., in prep.).

\begin{figure}
   \centering
   \includegraphics[width=9cm]{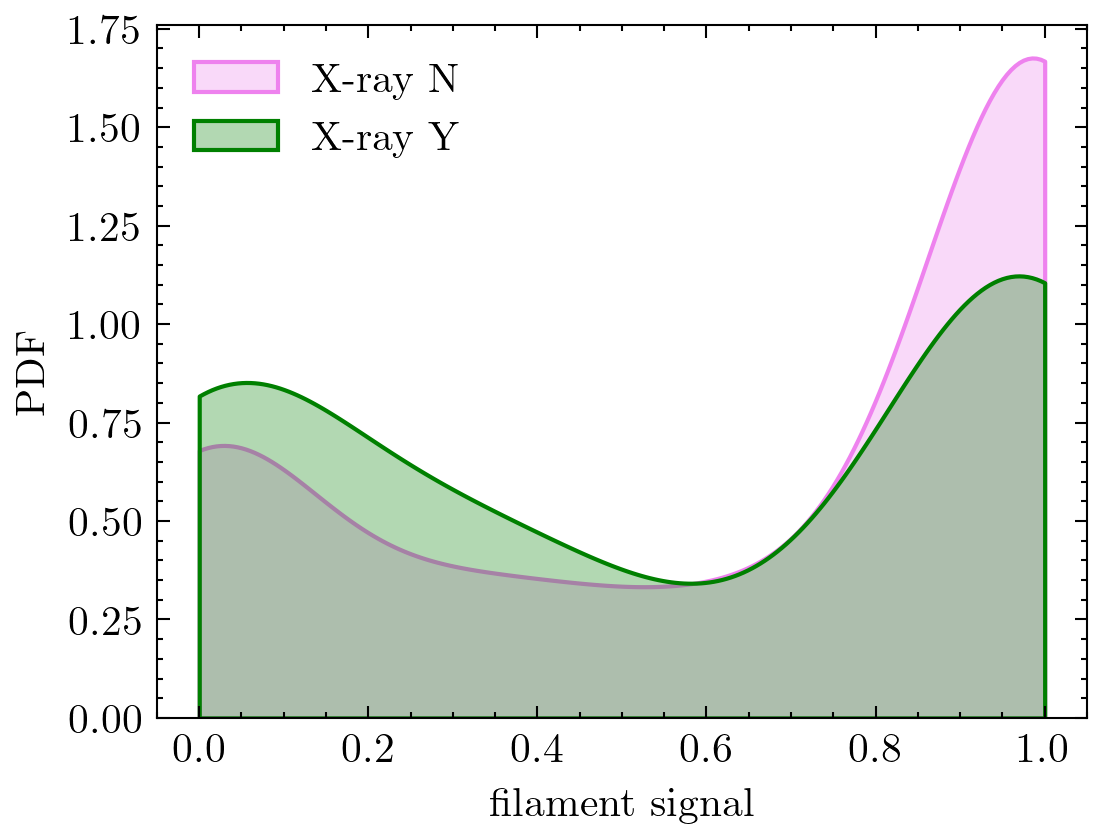}
      \caption{Filament signal (ratio of number of filament galaxies to number of filament and cluster galaxies) probability density function for X-ray bright ($Y$) and X-ray faint ($N$) groups, using the LSS data and galaxy classification by Taamoli et al., in prep. Consistent results were obtained using the LSS information provided by \citet{darvish_cosmic_2014}.}
         \label{lss}
\end{figure}

The results of this test are shown in Fig. \ref{lss} for the COSMOS2020-based LSS; results for the LSS map by \citet{darvish_cosmic_2014} show very similar trends. In Fig. \ref{lss}, the horizontal axis represents the filament signal, while the vertical axis indicates the probability density distribution smoothed and normalized using Kernel Density Estimation (KDE). To compute the filament signal, we counted galaxies in the central 0.25 Mpc/$h$ of each group and defined the filament signal as the ratio of the number of galaxies identified as part of a filament to the total (number of filament galaxies + number of cluster/node galaxies). In this way, groups with a filament signal close to 0 are likely hosted by a node (they are rich in cluster/node galaxies), while groups with filament signals close to 1 sit further away from nodes, likely found in filamentary environments connecting nodes. Groups without significant X-ray emission (pink curve) have around 2 times higher density in filaments (filament signal $>$ 0.8) with respect to X-ray bright ones (green curve). The latter are instead more likely to be found in denser environments in correspondence or close to cosmic-web nodes (filament signal $<$ 0.2).

Interesting insights can be gained from the investigation of group properties through a comparison between optically selected and X-ray selected clusters and groups. A comprehensive analysis of this aspect, for what concerns the AMICO-COSMOS groups and that incorporates the X-ray characterization of the new COSMOS-Web group catalog, will be presented in a forthcoming dedicated paper.

\section{Red sequence}\label{redseq}
We now turn our attention to how the RS in our group sample evolves over a wide range of redshifts, from $z=0$ to $z=3.7$. To study this, we need to identify the RS and characterize it through the main parameters of the RS ridgeline. 

\subsection{Observed red sequence with rest-frame matching
}\label{obsrs}
The first step to robustly detect the observed RS is to identify the appropriate combination of colors and magnitude that, at a given redshift, allows a clean identification of the sequence in the CMD. This can be done following the matched rest-frame photometry method \citep{blakeslee_clusters_2006, stott_evolution_2009}, namely by choosing two magnitude bands that bracket a characterizing spectral feature, like the 4000 $\AA$ break, and the reddest of the two as the reference magnitude for the CMD. This reduces the impact of the $k$ correction and the redshift dependence of the colors, straddling a feature that strongly traces different stellar populations in RS and non-RS galaxies.

\begin{figure}
   \centering
   \includegraphics[width=9cm]{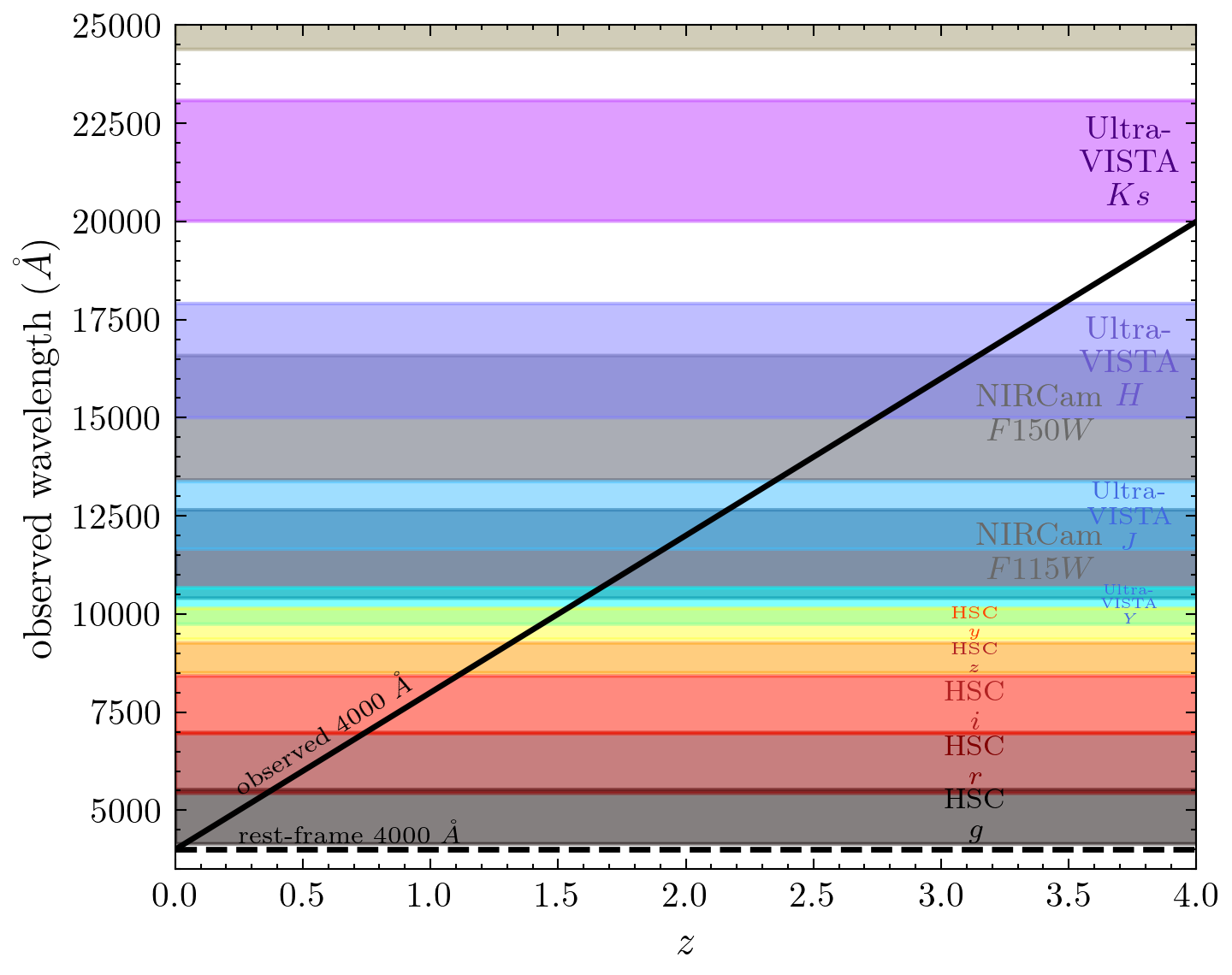}
      \caption{Sampling in the COSMOS-Web filters of the observed 4000 $\AA$, the spectral feature used for RS detection (solid black line). Different colors indicate different bands, as in the labels in the plot. Pairs of bands bracketing the break are reported in Table \ref{tab:redseq_colors}. The complementarity of COSMOS-Web dataset filters allows for a contiguous coverage up to $z\sim 3.5$.}
         \label{4000A}
\end{figure}

The COSMOS-Web galaxy catalog features a comprehensive set of magnitude bands spanning from optical to mid-infrared wavelengths. This coverage is uniform and continuous thanks to the complementarity of the JWST NIRCam filters and the UltraVISTA filters, making this the perfect opportunity to study observed RS evolution over a wide interval of redshifts, using the matched rest-frame photometry technique. This can be seen in Fig. \ref{4000A}, which shows the observed wavelength of the rest-frame 4000 $\AA$ break (solid black line) at different redshifts over the studied interval, and which filters are sampling and bracketing it. As clearly visible, the filter coverage of this spectral feature is continuous until at least $z \sim 3.5$, namely virtually covering the full COSMOS-Web group sample. Our selection of the 4000$\AA$-bracketing band pairs and the relative intervals are shown in Table \ref{tab:redseq_colors}. 
Once the RS is sampled in consistent observed colors, the ridgeline can be identified as the best-fitting line representing the locus of the quiescent member galaxies in the CMD.

\begin{table}[ht]
\centering
\caption{Red sequence selection method: observed color and magnitude choices by redshift interval.}
\begin{tabular}{ccc}
\hline
Redshift Range & Color & Magnitude \\
\hline
$0.01 < z < 0.38$ & $g - r$ & $r$ \\
$0.38 < z < 0.70$ & $r - i$ & $i$ \\
$0.70 < z < 1.10$ & $i - z$ & $z$ \\
$1.10 < z < 1.40$ & $z - Y$ & $Y$ \\
$1.40 < z < 1.70$ & $Y - J$ & $J$ \\
$1.70 < z < 2.35$ & $\mathrm{F115W} - \mathrm{F150W}$ & $\mathrm{F150W}$ \\
$2.35 < z < 3.25$ & $\mathrm{F150W} - H$ & $H$ \\
$3.25 < z < 3.70$ & $H - Ks$ & $Ks$\\
\hline
\end{tabular}
\label{tab:redseq_colors}
\end{table}

\noindent As in Sect. \ref{fredsec}, we performed this RS analysis over 4 magnitudes from the $m_\star(z)$ in the F150W band, which, given the depth of the galaxy catalog, guarantees redshift-dependent completeness until at least $z\sim 2.5$, without compromising the statistics. We computed the RS parameters by performing a weighted least squares (WLS) fitting analysis of the group CMDs on all galaxies attributed to a group with a probability larger than the minimum threshold imposed by AMICO; that is, 0.005. In a similar way to what we have done for the red fraction in Sect. \ref{basic_fred}, we weighted each galaxy with the probability of being quiescent, conditional on the probability of being a group member. This represents the first guess for the position of the RS, which is nevertheless still potentially contaminated by central and/or bright star-forming galaxies that might draw the ridgeline guess toward bluer colors. To refine the position guess, we performed an iterative cleaning of outliers by fitting with a 3$\sigma$ clipping. Once convergence is reached (over a maximum of five iterations), the following RS best-fit parameters are stored for each group: the scatter computed on final retained galaxies, the average color, which is the color of the RS ridgeline computed at $m_\star(z)$ and the number of RS galaxies, which we defined as the number of quiescent member galaxies ($P_{red}$ > 50\% and $P$ > 
50\%) lying within a typical $\pm 0.3$ scatter \citep[e.g.,][]{de_lucia_build-up_2007, martinet_evolution_2015} from the best-fit line.

\begin{figure}
   \centering
   \includegraphics[width=9cm]{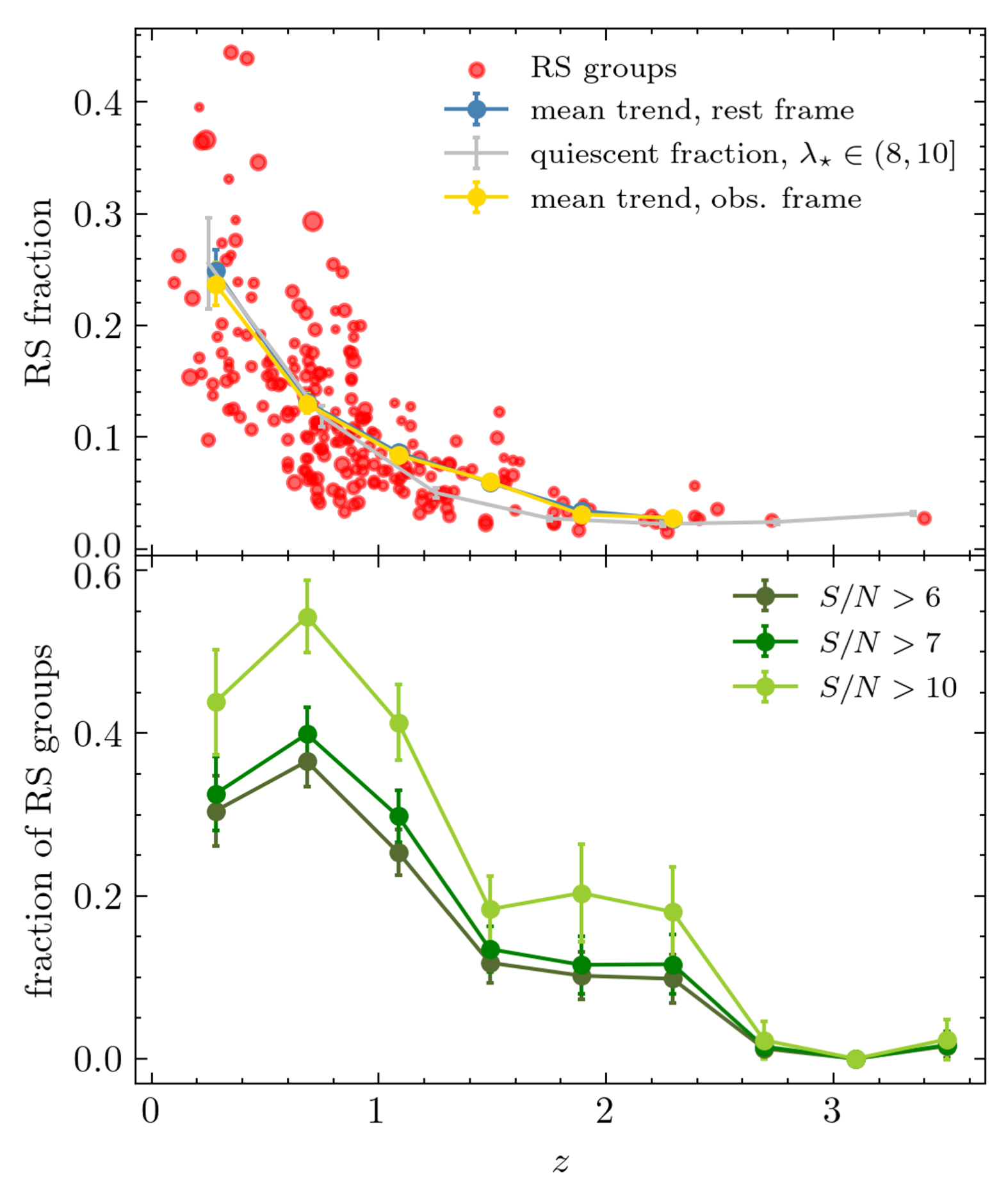}
      \caption{Top panel: Fraction of the selected RS galaxies over the total ($\lambda_{RS}/\lambda$) in each group, as a function of redshift for the sample with identified RS. The yellow and blue lines indicate the mean fraction in each $z$-bin weighted by the group $S/N$ (size of red points is proportional to the group $S/N$), using matched observed and rest-frame colors for the fit, respectively. Bottom panel: Fraction of groups with identified RS at different $S/N$ (or purity) levels, as a function of redshift, showing the first consistent appearance of RS in groups at $z\sim 2-2.5$.}
         \label{rs_fraq}
\end{figure}

Besides the direct count of RS galaxies, we also computed $\lambda_{RS}$, which is the sum of the membership probability of these RS galaxies, in analogy to the AMICO richness. Figure \ref{rs_fraq} shows the fraction of $\lambda_{RS}$ to the $\lambda$ (sum of all membership probabilities of galaxies used for the initial guess) as a function of redshift, for the sample of groups with an RS identified with $\lambda_{RS}>2$. This is an arbitrary limit that roughly corresponds to having at least 3 quiescent galaxies on the RS ridgeline, and therefore to a robust RS detection. The yellow line and points indicate the average trend of the RS fraction with redshift for the full sample. The size of the red points is proportional to the $S/N$ of the group detection, which is used as a weight for the average. The gray trend line is reported from the quiescent fraction analysis for the central richness bin (see Fig. \ref{basic_fred_fig}), while the blue trend line shows the same as done with the apparent magnitude but using rest-frame $J$ and $Ks$, which we describe in Sect. \ref{rf}. Quiescent fractions and RS fractions differ in terms of how they are defined and computed. The former is based on the probability of a galaxy being a member and being quiescent, while the latter also leverages the color homogeneity of the quiescent population (extracted by fitting a tight relation in the CMD). We find consistent trends between the two independent computations. 

In the bottom plot of Fig. \ref{rs_fraq}, we show the fraction of groups with an identified RS with $\lambda_{RS}>2$ over the total number of detections per redshift bin, for $S/N$-cut at 6, 7, and 10 (i.e., purity level of 77\%, 80\%, and 90\%). The overall trend follows that of the quiescent fraction and RS fraction, with the buildup of the RS settling between $z=2$ and $z=2.5$. The peak at $z \sim 0.7-0.8$ may be due to the presence of the COSMOS Wall \citep{iovino_high_2016}, which hosts more mature systems with established RSs. The total number of groups with an identified RS in this way is 214. 

The highest-redshift RS detected in this study is located at $z=3.4$ and it contains three quiescent member galaxies with very similar colors within around 250 kpc/$h$ of each other. This high-$z$ object is among the new protocluster cores not known in the literature before, which we recently discovered \citep{toni_cosmos-web_2025}\footnote{and detected independently in a parallel work presented by \citet{hung_discovering_2025}}. The protocluster core is rich ($\lambda_\star \sim 25$, $\lambda \sim 119$) and contains five galaxies with spectroscopic redshift, according to our spectroscopic counterpart assignment in the COSMOS spec-$z$ compilation \citep{khostovan_cosmos_2025}. The spectroscopic redshifts are compatible with that assigned by AMICO, solely based on photometry, which confirms the presence of an overdensity of galaxies at this location. However, none of the quiescent galaxies of the RS we found have a public spec-$z$ available. If confirmed, this would represent one of the highest-redshift examples of early RS detected to date (see the $z \sim 4$ protocluster reported by \citeauthor{tanaka_protocluster_2024} \citeyear{tanaka_protocluster_2024}, and the overdensity discovered at $z=3.44$ by \citeauthor{jin_cosmic_2024} \citeyear{jin_cosmic_2024}). The CMD of this $z=3.4$ detection (ID CW117) is shown in Fig. \ref{highzrs}, where we plot AMICO member galaxies with size proportional to their membership probability and their color referring to the probability of being quiescent and field galaxies as gray points. In the plot, we also report the main parameters of the RS ridgeline and the RS richness ($\lambda_{RS}$), while the RS galaxies are marked by black circles around the points. The dashed orange line indicates the expected color of a population passively evolving since $z=5$, which is our reference model. This represents the color of a typical luminosity-function-knee galaxy (with magnitude $m_\star$), according to the same configuration chosen for the AMICO model used in the group search, and mentioned earlier. The model was computed with the evolutionary synthesis model generator GALEV \citep{kotulla_galev_2009}, relying on a  Kroupa initial mass function \citep{kroupa_initial_2002} and adopting a chemically consistent approach \citep[see][for further details]{kotulla_galev_2009} for a massive elliptical formed at $z_f=8$ with a more recent medium-intensity star formation burst at $z_b=5$. This approach is similar to the one already used by \cite{castignani_star-forming_2022,castignani_star-forming_2023}. In the bottom left of the panel, we report the sky distribution of the galaxies in the central region of this high-$z$ detection, with the same notation as in the CMD plot.

\subsection{Results and comparison with rest-frame red sequence}\label{rf}
We performed RS fitting with the same procedure described above (3$\sigma$-clipped fit on WLS initial guess) based on the \texttt{LePhare} rest-frame magnitudes in the UltraVISTA-$J$ and -$Ks$ bands available in the COSMOS2025 release. We confirmed all RS detected with the observed magnitudes, mostly with compatible parameters. We additionally detected RS (mainly with low richness, $\lambda_{RS} \leq 3$) for another 11 groups.

\begin{figure}
   \centering
   \includegraphics[width=9cm]{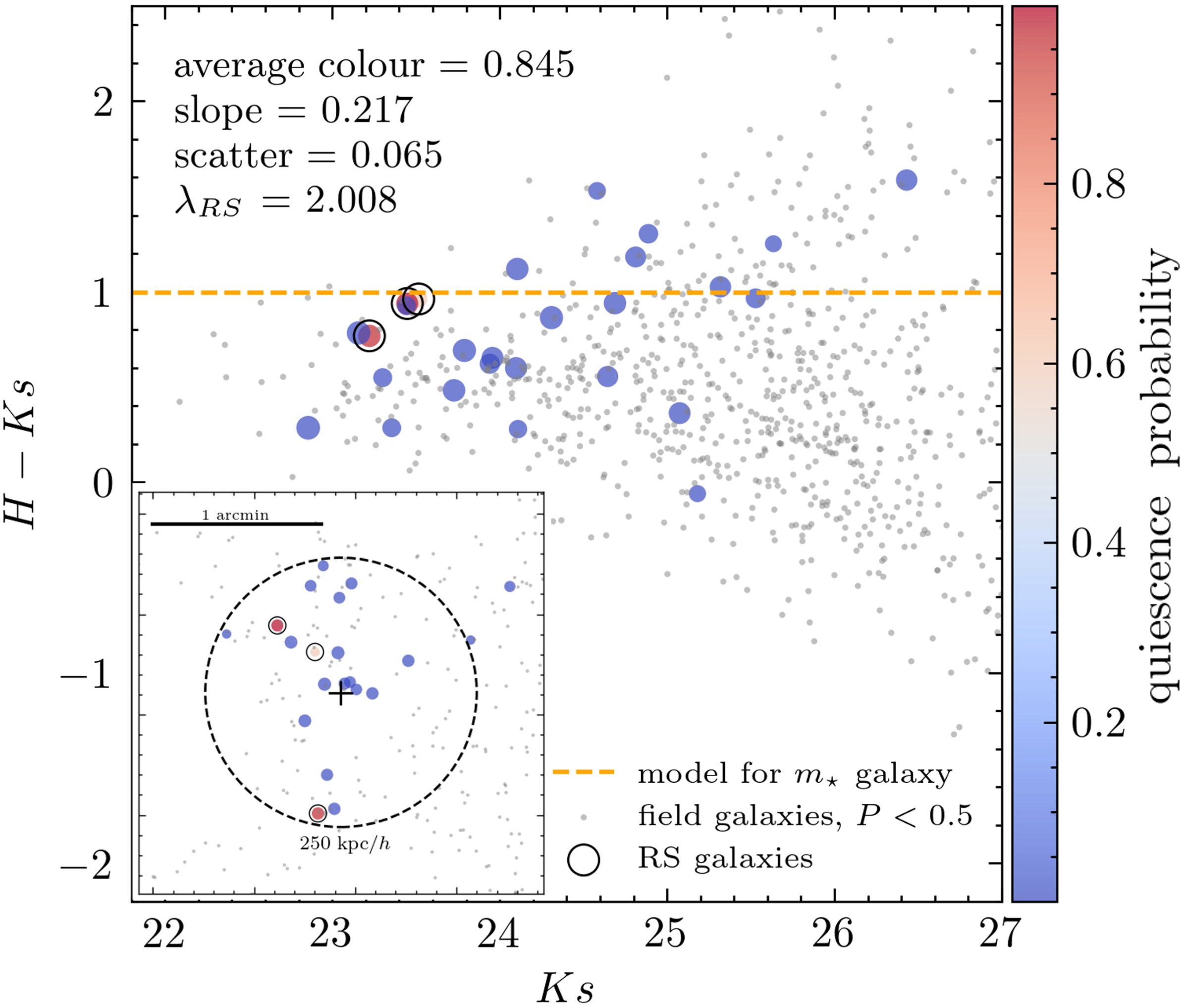}
      \caption{Observed-frame CMD of a rich protocluster core at $z=3.4$ (CW117) with three quiescent galaxies (red points with black contours) with colors consistent with RS evolutionary synthesis models for a typical $m_\star$ elliptical (orange line). This is the highest-$z$ RS we detected, which, once confirmed, would represent one of the earliest hints of RS ever discovered.}
         \label{highzrs}
\end{figure}

\begin{figure}
   \centering
   \includegraphics[width=9cm]{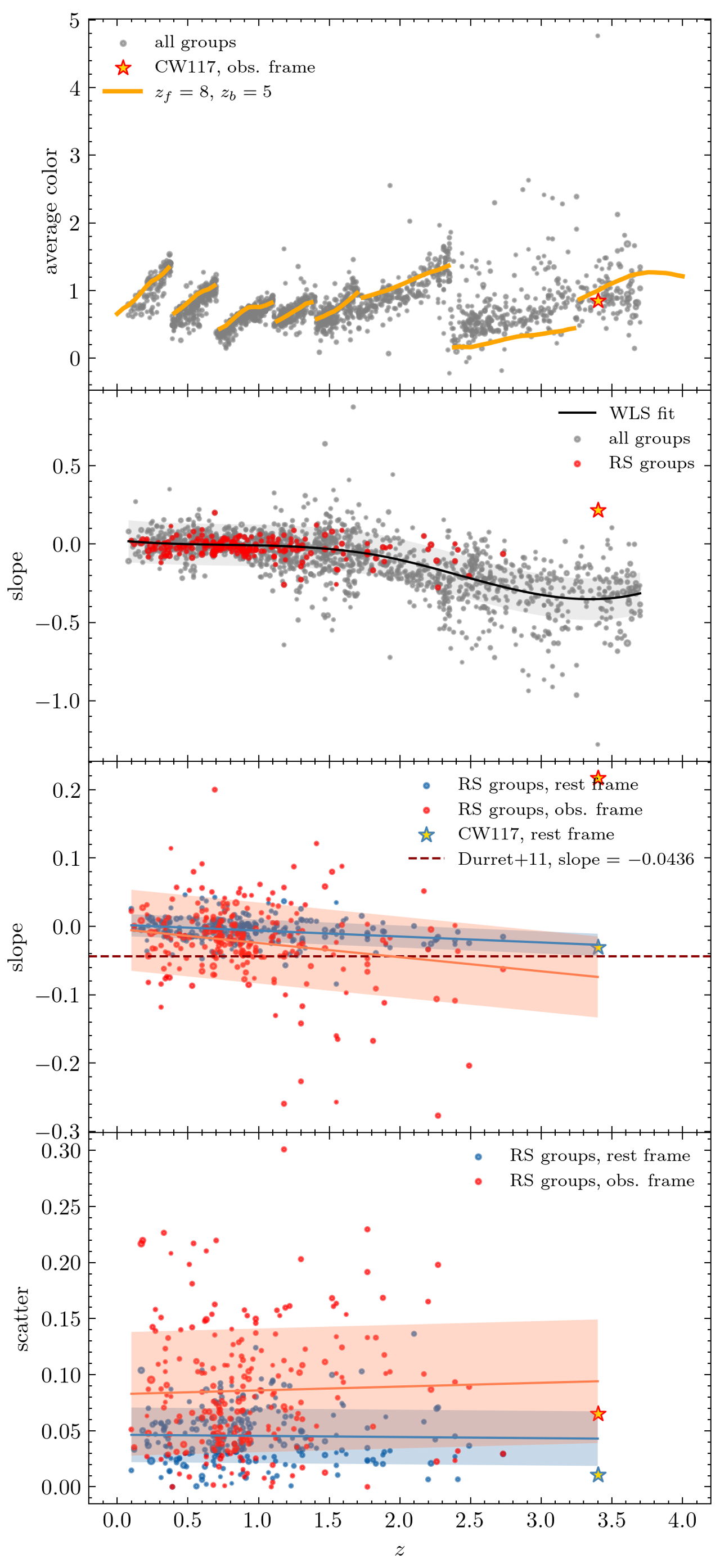}
      \caption{Evolution with redshift of the main RS parameters for our group sample. From top to bottom: the first panel shows the average observed color as in Table \ref{tab:redseq_colors}, compared to our reference $m_\star(z)$ model (orange lines). The star shows the CW117 detection, which has RS color compatible with that expected at that redshift for typical RS galaxies; the second panel shows the slope, with all groups in gray and those with RS in red; the third panel is a zoom on the slope of RS-groups only, with the parameters as retrieved using observer-frame (red points and orange fit line) and rest-frame magnitudes (blue point and fit line); the fourth panel shows the same but for the scatter of RS-groups. The yellow star shows the RS parameters of CW117, in the observer frame (red contours) and the rest frame (blue contours).}
         \label{rs_param}
\end{figure}

The RS parameters as a function of redshift for the RS ridgelines detected in this study are shown in Fig. \ref{rs_param}. When computing the fitted relations and mean trends, each group candidate is weighted by its $S/N$ to suppress the impact of the smallest and least reliable detections on the overall fit. The size of the points is proportional to the $S/N$ of the groups. The top panel shows the average observed color for all the detections in the COSMOS-Web group catalog, compared to the reference model we described above in Sect \ref{obsrs} (orange lines). Colors are computed in the associated redshift-dependent bands, as summarized in Table \ref{tab:redseq_colors}, which explains the jumps in average color when the 4000 $\AA$ switches between adjacent bands. Besides this effect, the trend of the model is consistent with the evolution of the observed colors in our sample until at least $z\sim 2.4$. For the last two redshift intervals, the observed color of the ridgeline is not consistent with a prototypical quiescent galaxy, which is expected since at this redshift we do not fit any RS for most of the groups, and the ridgeline parameters are tentative and mainly driven by bright star-forming galaxies populating groups and cores at such high redshifts. The location in the plot of the highest-$z$ RS presented above (CW117) is indicated by a yellow star. As already mentioned, the average observed color of this RS is consistent with the model. 

The second and third panels show the ridgeline slope. In the second plot, we include all groups in the sample (gray points) with the relative total fitted trend (black line) and its 1$\sigma$ error (shaded area). Red points mark groups for which it was possible to identify a RS with $\lambda_{RS}>2$. This latter sub-sample is shown in the third plot from the top, where we compare the slope computed with observed rest-frame-matched colors (red points and orange line and shaded area) with that based on rest-frame $J$ and $Ks$ magnitudes (blue points and blue line and shaded area). The two trends are consistent within confidence intervals, with the rest-frame colors offering a stronger constraining power. Under the assumption of negligible $k$-correction effect, the RS slope trend reflects metallicity evolution \citep[e.g.,][]{stott_evolution_2009}. We found a slight increase in the absolute value of the slope with redshift. However, the steepening of the RS over the studied redshift range is very mild, and there is no clear evidence of a significant evolution. The average RS slope evolves of only $\Delta \mathrm{slope} \sim 0.05$ across 12 Gyrs. This weak trend is consistent with little or no evolution in the mass–metallicity relation of quiescent galaxies, in agreement with, for instance, what was found for clusters up to $z\sim 1.5$ by \citet{mei_evolution_2009} and \citet{cerulo_accelerated_2016}. Therefore, our results support this picture, indicating that the mass–metallicity relation was already in place and has remained largely preserved since even earlier cosmic times. A slope for the RS typically adopted in the literature \citep{durret_merging_2011, martinet_evolution_2015, takey_3xmmsdss_2019} is shown as a dashed dark red line for comparison. Once again, the location of CW117 is indicated by a yellow star, with red contours for observed colors and blue contours for the rest frame. The slope of the RS in CW117 is a clear outlier in the observed colors (red star in the second and third panels of Fig. \ref{rs_param}), because the quiescent galaxies lie very close together in the CMD (see Fig. \ref{highzrs}), which drives the fitted slope to become large and positive.

As for the slope, our results for the RS scatter (fourth panel of Fig. \ref{rs_param}) show little to no evolution over the studied redshift range. The RS scatter is found to be below 0.1 mag for most of the detections ($\sim$64\%), especially when measured in the rest frame ($\sim$98\%). The persistence of small scatters might suggest that the bulk of stars in RS galaxies formed early and over relatively short timescales, with limited diversity in their subsequent star formation histories. However, a more quantitative interpretation, such as estimating age dispersion, depends sensitively on photometric uncertainties, color choice, RS definition, group assembly history, luminosity thresholds, and survey depth; therefore, it is not straightforward to draw firm conclusions about the absolute age spread of the population from the observed RS scatter. However, the average scatter values and trend behavior are consistent with previous studies based on both simulations and semi-analytical models \citep[e.g.,][]{menci_red_2008,romeo_study_2015} and on observations \citep[e.g.,][]{hao_precision_2009,mei_evolution_2009,cerulo_accelerated_2016}. We found that the scatter little-to-no evolution reported by \citet{cerulo_accelerated_2016} for clusters up to $z=1.5$ persists at higher redshifts, with an average rest-frame value of $\approx 0.045 \pm 0.025 $ mag, consistent with that found by \citet{mei_evolution_2009} for elliptical ($0.042 \pm 0.021$ mag) and for elliptical + lenticular galaxies ($0.062 \pm 0.015$ mag). When considering observed colors, average RS scatter and its spread increase with respect to rest-frame measurements ($\approx 0.084 \pm 0.054$), consistent with the low-redshift observations by \citet{hao_precision_2009} and semi-analytical model predictions \citep{menci_red_2008}.

\section{Conclusions}\label{conclusions}
We conducted a study of the evolution in redshift and the dependence on group richness of the quiescent galaxy fraction and of the RS parameters for COSMOS groups detected with the AMICO algorithm, across around 12 Gyrs of cosmic history, from $z=0$ to $z=3.7$. We used multiple complementary methods to estimate the quiescent fraction and detect the RS, finding consistent results across different approaches. In total, we identified 225 RS ridgelines using rest-frame colors, with 214 also consistently characterized using rest-frame-matched observed colors. 
The main results of our study are as follows:
   \begin{itemize}
      \item Machine learning classifiers (both linear and nonlinear) provide a powerful framework for leveraging photometric multiband survey data to robustly distinguish between quiescent and star-forming galaxies. In particular, gradient boosting decision trees, such as \texttt{XGB}, offer efficient handling of missing data and class imbalance, while delivering reliable probabilistic classifications. In our application, when tested against a testing dataset, this approach achieved an F1 score exceeding 93\%, even in the case of limited availability of rest-frame magnitudes.
      \item The AMICO algorithm identifies galaxy groups without relying explicitly on color information. As a result, the AMICO catalogs enable a robust analysis of member galaxy properties, such as colors and luminosities, in both model-dependent and model-independent ways. Our results on the buildup of the quiescent galaxy population are found to be consistent across these methods.
      \item We found that the quiescent galaxy population in groups builds up steadily starting from $z = 1.5 - 2$ in all richness bins, with evidence of an earlier and accelerated growth for the richest systems, consistent with the scenario proposed, for instance, by \citet{cerulo_accelerated_2016}. 
      \item By analyzing X-ray characterized AMICO detection, we found an indication that X-ray faint groups show, on average, lower quiescent fractions than X-ray bright ones. This might be connected to their location in the cosmic web, as X-ray faint groups are more likely to be found in filaments, where the environmental pre-processing of galaxies is lower compared to cosmic nodes, as well as the local density.
      \item We detected and characterized 225 RS ridgelines in COSMOS-Web groups, providing an extensive redshift coverage. We report the discovery of a rare compact ($\sim 250$ kpc/$h$ scale) protocluster core with quiescent galaxies at $z=3.4$, which, if confirmed, may represent one of the most distant early RSs detected to date, not known in the literature before our recent group search in COSMOS-Web.
      \item We found that the observed average colors of the RS ridgelines in our sample are consistent with those predicted by a passively evolving elliptical galaxy model. Regarding the slope and scatter of the RS, we observed no significant evolution with redshift, suggesting that the absence of significant trends found in previous studies of more massive galaxy clusters and at lower redshift may persist in the group regime and at earlier epochs.
   \end{itemize}

\noindent This study extends the detection and characterization of the RS and quiescent galaxy fractions beyond $z>1.5$ and into the regime of galaxy groups. Such an analysis is a natural application of a deep and well-characterized dataset such as the COSMOS-Web group catalog. Especially for high-$z$ candidates and RS evolution trends, we stress the necessity of spectroscopic confirmation, which we plan to carry out in a dedicated campaign. In forthcoming work, we shall further investigate the correlation between galaxy properties and X-ray emission (i.e., intracluster and intragroup medium properties) and its connection with the LSS. Additionally, we plan to explore the properties and role of the central brightest group galaxies, examining how their evolution relates to the environment and the assembly history of their hosts.

\begin{acknowledgements}
      The authors thank the anonymous referee for the careful reading and the very useful suggestions, which helped improve the robustness and clarity of this manuscript. GT thanks David P. Petri for the valuable discussion and useful suggestions regarding ML. LM acknowledges the financial contribution from the PRIN-MUR 2022 20227RNLY3 grant “The concordance cosmological model: stress-tests with galaxy clusters” supported by Next Generation EU and from the grant ASI n. 2024-10-HH.0 “Attività scientifiche per la missione Euclid-fase E".
\end{acknowledgements}

\bibliographystyle{aa} 
\bibliography{references_merged}

\begin{appendix}
\section{Training-set choice and labeling}\label{appendixA}

\subsection{Additional details about the training-set choice}

Here we provide additional details on the selection and use of the training set, highlighting its strengths and limitations in the context of our analysis. Firstly, we chose not to train our model directly on the target dataset because doing so would require extracting data from it and splitting it into training and testing subsets, thereby drastically reducing the statistical completeness of the final catalog that we aim to fully analyze. Instead, we train the model on an independent labeled dataset (COSMOS2015) while applying it to the full target sample (COSMOS2025). Using real data from the same field ensures that the model is exposed to the true photometric uncertainties, systematics, blending issues, and observational conditions of our survey, which simulations might fail to fully reproduce.

COSMOS2025 lies entirely within the COSMOS2015 footprint, so some galaxies appear in both catalogs. However, the two surveys differ in instrument sets, depths, PSFs, and photometric methods, resulting in different galaxy characterization, including different rest-frame magnitudes. From the ML perspective, what matters is overlap in feature space, not sky coordinates. We verified that no galaxy has identical or effectively duplicate feature vectors in both datasets, meaning there is no risk of data leakage or overfitting. COSMOS2015 thus provides an independent, realistic training set for our application to COSMOS2025.

The redshift range of the training set extends to $z=6$, while the target set reaches $z\sim13$. However, galaxies with $z>6$ constitute only 0.6\% of the target sample and are negligible for our analysis. The main gap in the training set we chose in our approach lies instead in the depth mismatch: the training set is limited to objects brighter than $H =25$~mag, while the target set extends to $27.3$~mag. This forces the model to extrapolate beyond its training domain, where faint objects have larger photometric uncertainties and somewhat broader color distributions. Consequently, classification performance may degrade at the faint end, potentially increasing systematic misclassifications.

To assess the practical impact, we examined galaxies fainter than $H=25$ in the classified target set, using the widely adopted NUVrJ diagram as a reference. We analyzed (1) the distribution of predicted probabilities and (2) classification metrics in classifying quiescent galaxies\footnote{the performances for star-forming galaxies, which are the majority class, are not relevant here because they do not vary with magnitude, remaining stable above $\sim 96$\%.} (precision, recall, F1 score) for faint versus bright objects. The probability distribution for faint galaxies was found to be broadly consistent with that of the brighter sample, with a slight increase in intermediate probabilities ($P_{\mathrm{blue}}\sim 70$\%), reflecting model caution in ambiguous cases. Compared to the bright end, the faint sample shows increased precision ($\sim +10$\%) but substantially reduced recall ($\sim -30$\%), leading to a net modest decrease in the F1 score ($\sim -10$\%). This indicates that the classifier becomes more conservative at faint magnitudes: quiescence predictions are generally correct, but a larger fraction of quiescent galaxies are missed. In other words, the model favors purity over completeness in the unrepresented faint regime. We stress that this comparison measures consistency with NUVrJ rather than ground-truth validation, so it does not provide a direct correction for faint-end behavior.

Large uncertainties are in general expected in this regime, given the larger photometric errors at the faint end, and our model's conservative response is appropriate. Importantly, only $\sim 25$\% of group member galaxies fall in this faint regime (reducing to $\sim 14$\% when considering confident associations with $P_{\rm mem}>75$\%), limiting its impact on our main scientific goals. In conclusion, when deriving quiescent fractions, the faint end may be characterized by a selection that is purer but less complete, consistent with the intrinsic limitations of the data. Overall, the results indicate that the model effectively copes with fainter, noisier data, becoming more conservative, without compromising the analysis of galaxy population trends.

\subsection{Additional details about the training-set labeling thresholds}

Here we provide further information about the choice of labeling criteria and thresholds and their $z$-dependence.

The NUVrJ criterion classifies galaxies using rest-frame $NUV-r$ and $r-J$ colors, identifying as star-forming the galaxies with $NUV -r < 3(r - J) + 1$ or $NUV - r < 3.1$. The method was introduced by \citet{ilbert_mass_2013} as an alternative to the very commonly adopted UVJ diagnostic. It was developed as an alternative with extended validity up to high redshift, given its larger dynamical range, lower sensitivity to uncertainties, and better sampling at $z>2$. The NUVrJ criterion has been widely tested and employed for galaxy classification up to $z\sim 4$ and beyond \citep[e.g.,][]{ilbert_mass_2013,weaver_cosmos2020_2023,gould_cosmos2020_2023,shuntov_cosmos2025_2025,pearson_influence_2023}. 

Another variation that complements the NUVrJ criterion is the NUVrK threshold \citep[][]{arnouts_encoding_2013,davidzon_vimos_2016}, which replaces the $r-J$ with the $r-K$ color. Galaxies are classified as star-forming when $NUV - r < 1.37(r - K) + 2.6$ or $NUV - r < 3.15$ or $r - K > 1.3$. The NUVrK criterion, while with similar behavior and $z$-validity to the NUVrJ, combines the sensitivity of the $NUV-r$ color to the recent star formation history of the galaxy with the sensitivity of the $r-K$ color to stellar aging and dust attenuation, and helps also to better separate the influence of AGNs \citep[e.g.,][]{siudek_environment_2023,moutard_slow_2020,pearson_influence_2023}.

As for what concerns the sSFR cut, this has been widely used in the literature by adopting different thresholds, generally between $10^{-10}\,\mathrm{yr}^{-1}$ \citep[e.g.,][]{wu_stellar_2018} and $10^{-11}\,\mathrm{yr}^{-1}$ \citep[e.g.,][]{ilbert_mass_2013}. We decided to adopt an intermediate fixed cut at $10^{-10.5}\,\mathrm{yr}^{-1}$ which is the same adopted in Euclid \citep[][]{euclid_collaboration_humphrey_euclid_2023}, following the work of \citet{bisigello_euclid_2020}. However, it is true that the sSFR threshold to separate quiescent and star-forming galaxies is not uniform and can vary with cosmic time and with galaxy properties \citep[e.g.,][]{bisigello_euclid_2020, pacifici_evolution_2016,franx_structure_2008,pearson_influence_2023,wetzel_galaxy_2013}. 

\begin{figure}
   \centering
   \includegraphics[width=9cm]{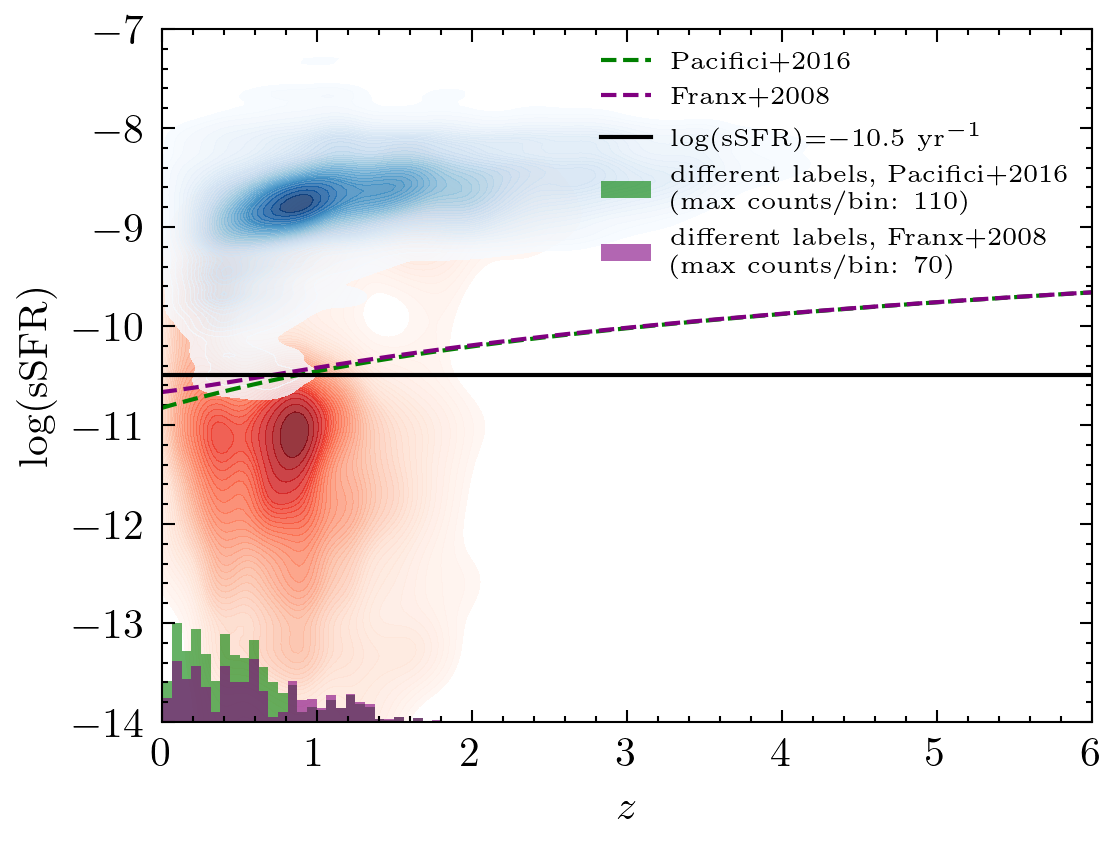}
      \caption{Density of COSMOS2015 star-forming (blue) and quiescent (red) galaxies (according to NUVrJ classification) in the sSFR–$z$ plane. The solid black line shows the fixed sSFR threshold used in this work; dashed lines indicate the time-dependent thresholds of \citealt{pacifici_evolution_2016} (green) and \citealt{franx_structure_2008} (purple). Thresholds diverge at $z>2$, but converge near $z\sim0.7$–$0.9$, where most COSMOS galaxies lie. The bottom histograms show the redshift distribution of galaxies that would change classification under the two time-dependent thresholds, as reported in the legend.}
         \label{pacfranx}
\end{figure}

In particular, \citet{franx_structure_2008} have proposed a threshold proportional to the Hubble parameter, $H(z)$, where quiescent galaxies have sSFR $< 0.3 H(z)$. In an analogous way, \citet{pacifici_evolution_2016} have more recently proposed a threshold dependent on the inverse of the age of the Universe, $t_U(z)$, with quiescent galaxies having sSFR $< 0.2/t_U(z)$. Figure \ref{pacfranx} shows the density of the selected COSMOS2015 star-forming (blue) and quiescent (red) galaxies according to the standard NUVrJ diagnostic in the sSFR vs. $z$ diagram. The solid black line indicates the fixed threshold in sSFR we applied in this work, while the dashed lines mark the time-dependent threshold suggested by \citet{pacifici_evolution_2016}, in green, and the one by \citet{franx_structure_2008}, in purple. The fixed and time-dependent thresholds diverge significantly at $z>2$, when low-sSFR galaxies are, however, expected to be much rarer. The thresholds become more consistent with each other around the crossing-point at $z\sim 0.7-0.9$ where most of the galaxies in COSMOS are \citep[e.g.,][]{iovino_high_2016}, as visible in the plot. At the bottom of the panel in Figure \ref{pacfranx}, we also show the distribution in redshift of all galaxies that would change their label if a time-dependent threshold was adopted. Counts for specific changes between the three classes of galaxies are also reported in Table \ref{change_label}. Two main conclusions can be drawn from these results: (1) the impact of a time-dependent sSFR is significant only below the crossing-points ($z \lesssim 0.7$) and affecting a small fraction of galaxies (overall $\sim$0.2$-$0.4\% of the galaxy sample); (2) this effect is mainly manifested in slightly reducing the number of quiescent galaxies ($\sim$ $1-2$\%), and the number of green valley galaxies ($\sim 1-8\%$) in favor of a few hundreds ($\sim$ $0.1-0.3$\%) of new examples of star-forming galaxies (especially when applying \citet{pacifici_evolution_2016}'s threshold). Given that the effect is limited in redshift, affecting only a small fraction of the galaxies, and that it reduces examples for the minority class (quiescent galaxies) and increases those for the majority class (star-forming galaxies) close to the decision boundary, it is not expected to drastically impact the learning process and performance.
In the redshift range we analyzed, the simple choice of $10^{-10.5}\,\mathrm{yr}^{-1}$ to separate the two populations is therefore well justified and closer to time-dependent relations with respect to, for instance a threshold at $10^{-11}\, \mathrm{yr}^{-1}$, which would largely overestimate the number of star-forming galaxies.

\begin{table}[ht]
\centering
\caption{Number of galaxies which would have been labeled differently if the threshold by \citet{franx_structure_2008} / \citet{pacifici_evolution_2016} was considered instead of a fixed threshold at $10^{-10.5}\, \mathrm{yr}^{-1}$. Rows indicate labels assigned in this study, and columns indicate new labels that would have been assigned with time-dependent thresholds.}\label{change_label}
\begin{tabular}{c|ccc}
\hline
& Quiescent & Star-forming & Green \\
\hline
Quiescent & -- & 255 / 458 &27 / 84 \\
Star-forming & 19 / 17 & -- & 204 / 123 \\
Green & 42 / 26 & 229 / 427 & -- \\
\hline
\end{tabular}
\end{table}

Finally, for the Sa-color evolution diagram \citep{andreon_build-up_2006,radovich_amico_2020}, a few more aspects have to be taken into account. The classification method is interesting because it uses apparent colors and not rest-frame colors. The use of an evolving Sa-model is what makes it redshift-consistent: each galaxy is classified as quiescent or star-forming, given the typical color of a Sa-type galaxy (a spiral galaxy with properties close to those of lenticular and elliptical galaxies according to Hubble's morphological classification scheme) at that specific redshift. However, the main limitation of this method is in the model-dependence itself and the fact that the chosen color must bracket the $4000 \AA$ break and therefore change accordingly as redshift increases. 

In this study, we modeled the evolution of Sa-colors with the GALEV evolutionary synthesis interface \citep{kotulla_galev_2009}, choosing typical Sa parameters and formation redshift. We decided to use this criterion only over a limited redshift range; that is, $0.38 \leq z \leq 1.4$. The lower bound is due to the absence of $g$-band photometry in the COSMOS2015 catalog caused by poor seeing \citep{laigle_cosmos2015_2016}, which would be the blue-bracket of the $4000 \AA$ break at $z<0.38$. The upper limit is instead given by the $4000 \AA$ break shifting into the near-infrared. Given that the uncertainties of our model increase in this regime, we chose to be conservative and have the ML algorithms rely only on the other three methods at higher redshifts. 

\begin{figure}
   \centering
   \includegraphics[width=9cm]{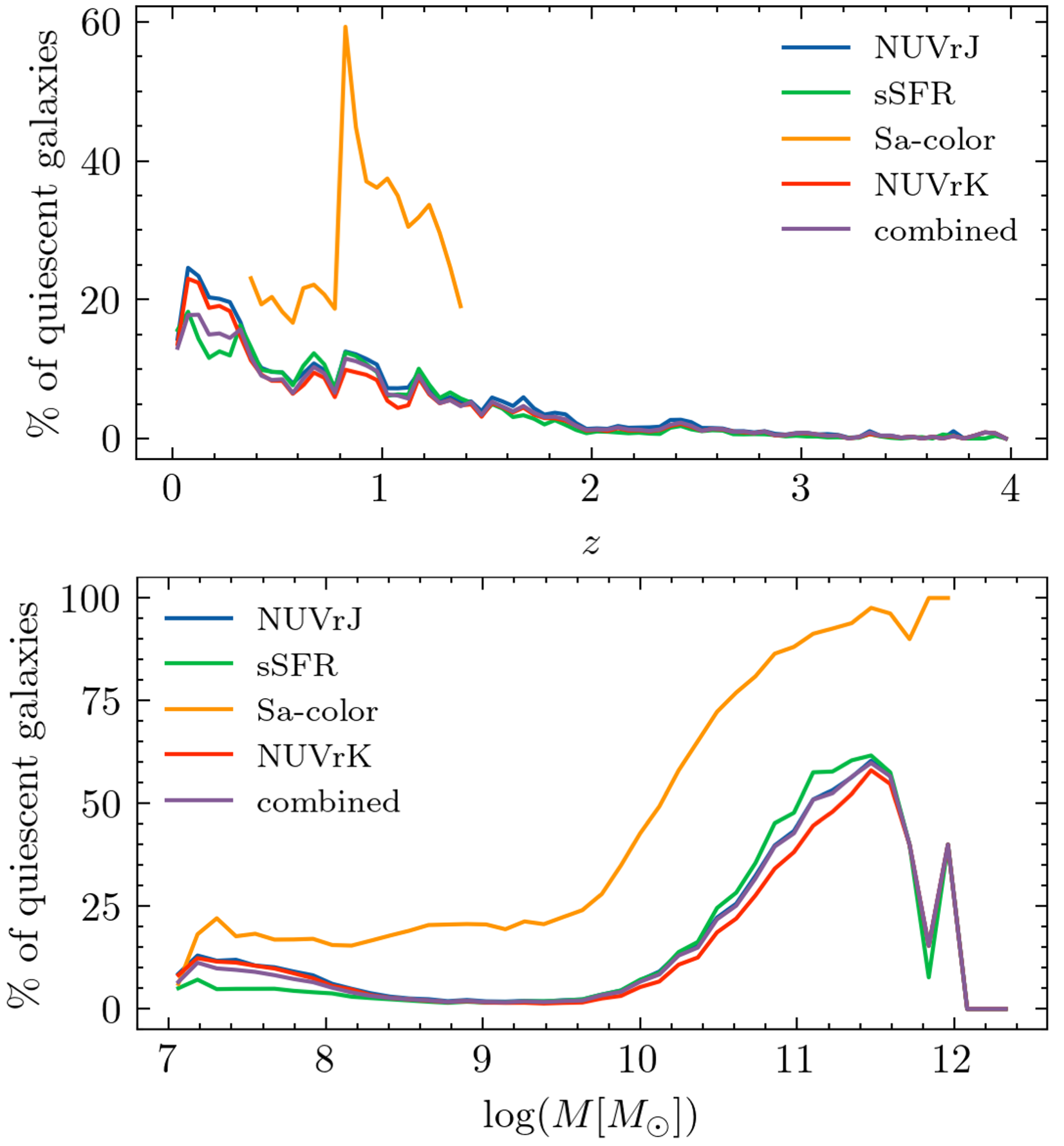}
      \caption{Variation in bins of redshift (upper panel) and galaxy stellar mass (lower panel) of the percentage of classified quiescent galaxies over the total, for the NUVrJ color-color cut (blue), the sSFR threshold (green), the Sa-color method (orange), and the NUVrK color-color cut (red). The purple line shows the final combined labeling we adopted for the training and testing sets.}
         \label{impact_criteria}
\end{figure}

Figure \ref{impact_criteria} shows the variation with redshift (upper panel) and with galaxy stellar mass (lower panel) of the percentage of classified quiescent galaxies over the total, for the four methods and for the final combined labeling we adopted (purple). As expected, even if quiescent galaxies are only 7\% of the overall sample, the percentage over the total is very redshift- and mass-dependent. The panels also clearly illustrate the strong consistency between the color–color cuts and the sSFR threshold across all redshifts and stellar masses, while highlighting a noticeable overestimate in the number of quiescent galaxies derived from the Sa-color method compared to the other approaches. The peak around $z \sim 0.8$ is likely influenced by a mismatch between our model and the observed colors when transitioning from the $i$ to the $z$ band. Apart from this effect, the number of quiescent galaxies remains systematically $15-20\%$ higher across nearly all redshifts and masses -- an excess over color-color cuts similar to that recently reported by \citet{asadi_machine_2025}. Interesting insights on this discrepancy could be gained by investigating a complete and spectroscopically characterized sample of green valley galaxies, which comprise most of the ambiguous and transitional cases in the classification.

In our specific application, as shown in Figure \ref{impact_criteria}, the resulting overall galaxy classification (purple line) is, by construction, driven by the consensus among methods and is therefore effectively constrained by the other three criteria. Nevertheless, the Sa-color method provides valuable information in ambiguous or edge cases and is particularly helpful when UV photometry is missing. Training the ML models on a labeled dataset that integrates multiple independent and complementary methods, while basing the labeling on their consensus, is one of the key strengths of our approach.

\section{Cylinder method and purity weighting of the quiescent fraction}\label{appendixB}
In Sect. \ref{cylsection}, we have introduced a new method to characterize AMICO galaxy group members that (nearly) eliminates the model-dependence (and therefore the dependence on magnitude and position), while retaining the information of the redshift probability distribution. In short, this method exploits the concept of considering a volume around each detection and subtracting the density of field galaxies at the corresponding redshift to remove field contamination. Two key requirements come with this formalism: the appropriate choice of the cylinder size and the handling of negative densities, since these may arise from the group–field subtraction.

The choice of the depth of the cylinder is motivated by the typical photometric redshift uncertainty in the redshift range of interest. The depth is variable and redshift dependent, set to $\pm 0.01(1+z)$, and allows for a consistent sampling of members along the line of sight. Since galaxies are weighted by their redshift probability distribution, the exact choice of depth is generally less constraining and less impactful than the choice of the sky radius.

\begin{figure}
   \centering
   \includegraphics[width=8cm]{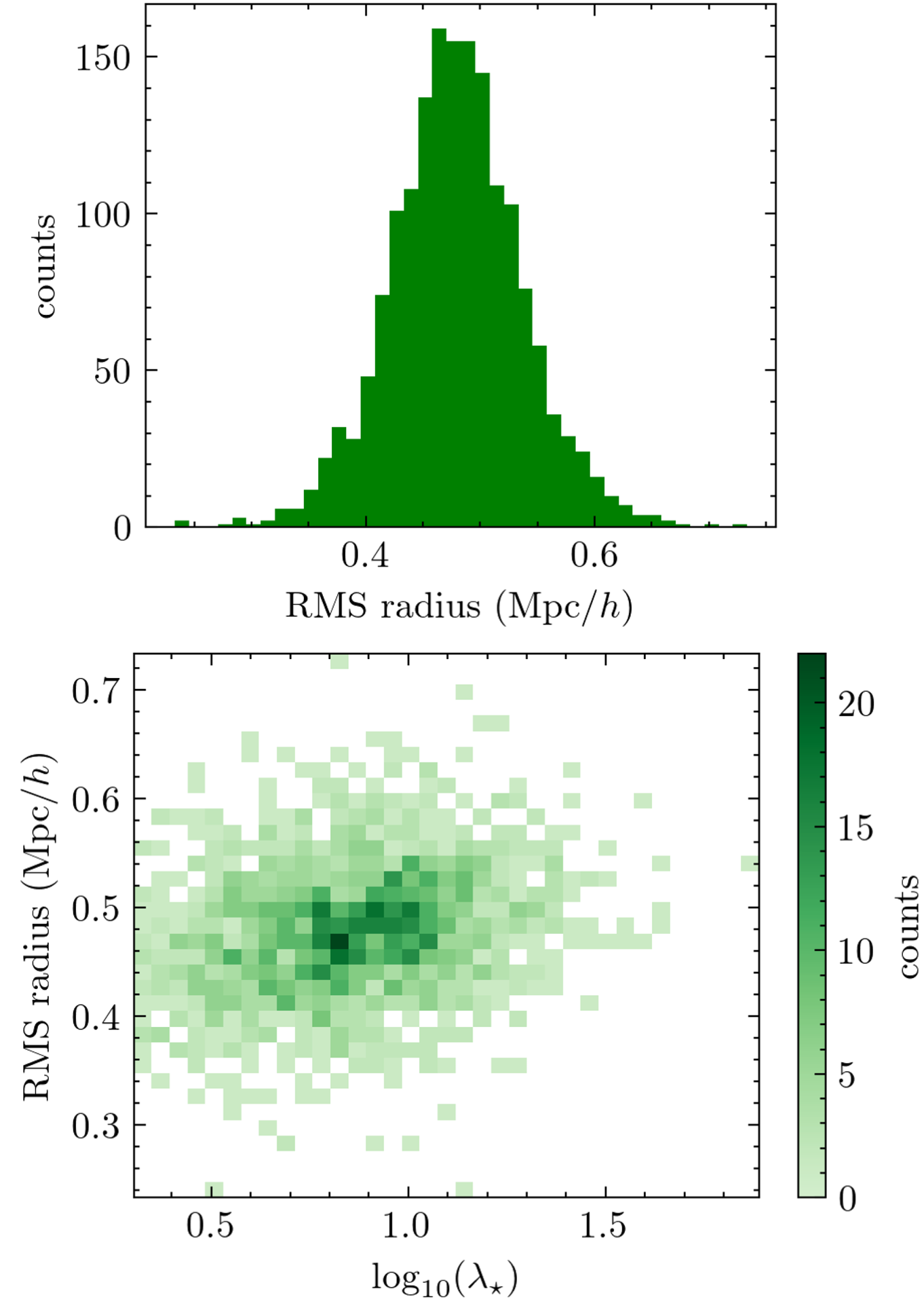}
      \caption{Top panel: distribution of RMS radii in Mpc/$h$ for the studied group sample. Bottom panel: trend of RMS radii with richness, log$_{10} (\lambda_\star$), shown as counts per bin.}
         \label{rms_radius}
\end{figure}

For choosing the cylinder sky radius, we started from an initial guess of $r = 0.5$~Mpc/$h$, and tested both $r$ and $2r$ as fixed cylinder radii. The differences in the resulting measurements and their richness dependence suggested that an adaptive radius, varying with the distribution of galaxies in each group -- and therefore tracing its richness and spatial extent -- would be a more robust and physically motivated choice. To determine this adaptive radius, we computed the RMS of all galaxies associated with a group, weighting each galaxy by its membership probability. The distribution of the resulting RMS radii and their relation to richness ($\lambda_\star$) are shown in Fig.~\ref{rms_radius}. Although the top panel indicates that 0.5~Mpc/$h$ was a reasonable initial guess, the adaptive radius consistently captures the group region across different group sizes, as demonstrated by its trend with richness (bottom panel).

As mentioned above, the subtraction of average field-galaxy densities from the group densities within the cylinder can theoretically produce negative densities (see Eq.~\ref{bgsred}). In other words, in the ratio of Eq.~\ref{final_ratio}: (1) the numerator, $\rho_\mathrm{red}$, can be negative (the density of red galaxies is higher outside the cylinder than inside, i.e., a particularly blue group); (2) the term $\rho_\mathrm{blue}$ can be negative (the density of blue galaxies is higher outside the cylinder than inside, i.e., a particularly red group); (3) the denominator can be negative overall (the group appears as an underdensity relative to the global field, but overdense with respect to the local field).

Cases (1) and (2), affecting a total of about 800 groups, are physically meaningful. We therefore flagged them and included them in our analysis by replacing negative densities with close-to-zero positive constants to avoid divergences. Case (3) affects only 10 low-richness groups (when using the adaptive RMS radius). If real, such systems might correspond to overdensities embedded within locally underdense regions, such as filament edges or voids. However, these detections are excluded from the present analysis, as the cylinder-based estimate cannot be determined reliably.

\begin{figure*}
   \centering
   \includegraphics[width=17cm]{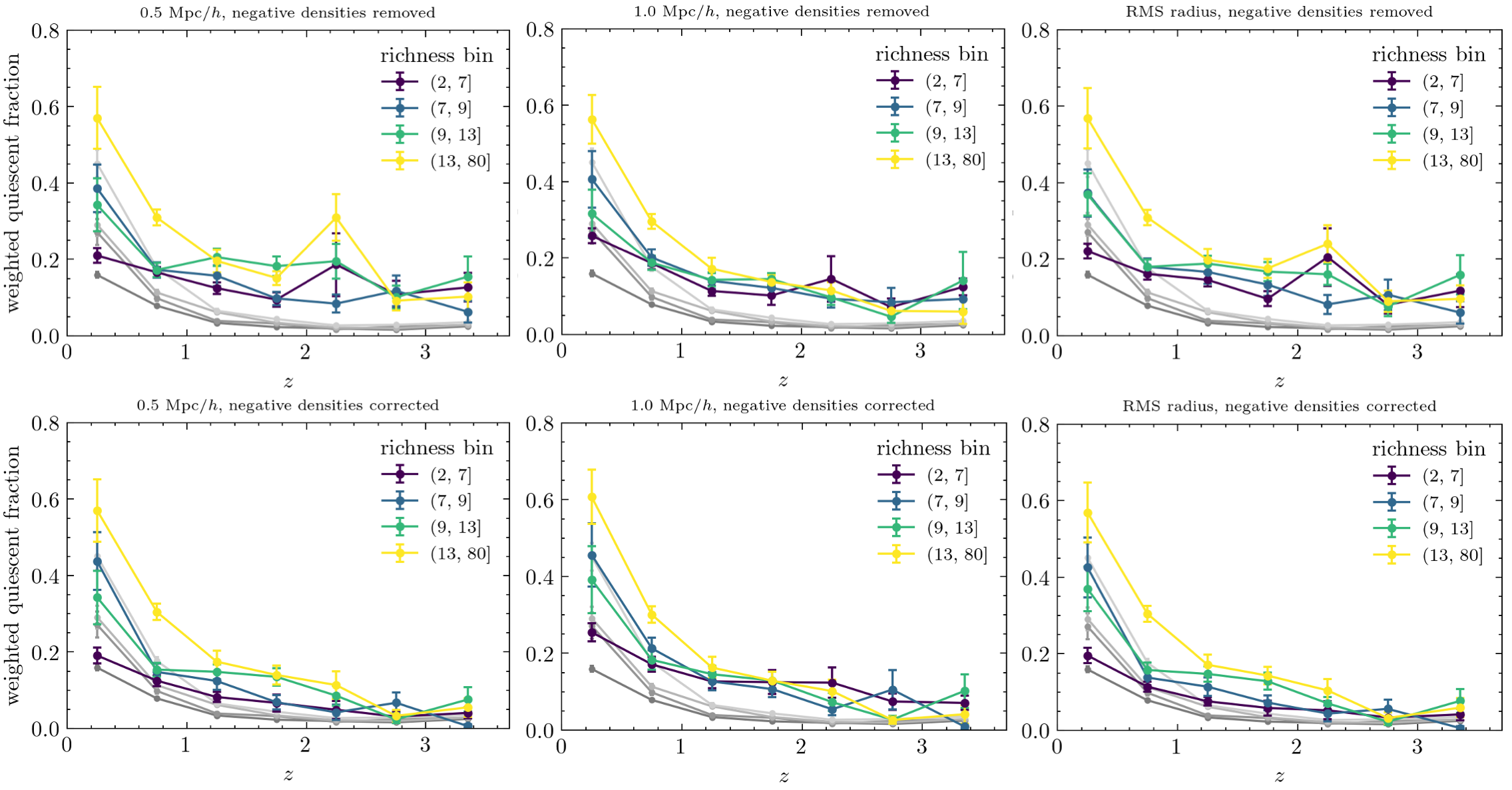}
      \caption{Purity-weighted quiescent fraction as a function of redshift for different richness bins, using fixed radii of 0.5 Mpc/$h$ (left), 1.0 Mpc/$h$ (middle), and the RMS radius (right). In the top panels, we discard all systems with negative densities, while in the bottom panels, we apply the corrections described in the text. Grayscale trend lines are reported using the pure-membership method in the same bins, for comparison. The RMS-based measurement with corrected negative densities is the method adopted in Sect. \ref{cylsection} and shown in Fig.~\ref{fred_cyl}.
}
         \label{cylinder_panel}
\end{figure*}

Figure~\ref{cylinder_panel} shows the weighted quiescent fraction as a function of redshift for different richness bins using 0.5~Mpc/$h$ (left column), 1.0~Mpc/$h$ (middle column), and the adaptive RMS radius (right column), comparing the results obtained by discarding all groups with any case of negative densities (top row) to those obtained after correcting negative densities in cases (1) and (2) as described above (bottom row). The RMS-based quiescent fraction with these corrections is the final method adopted for the results presented and discussed in Sect.~\ref{cylsection} and shown in Fig.~\ref{fred_cyl}.

\subsection{Quiescent fraction purity-weighting}
\begin{figure}
   \centering
   \includegraphics[width=9cm]{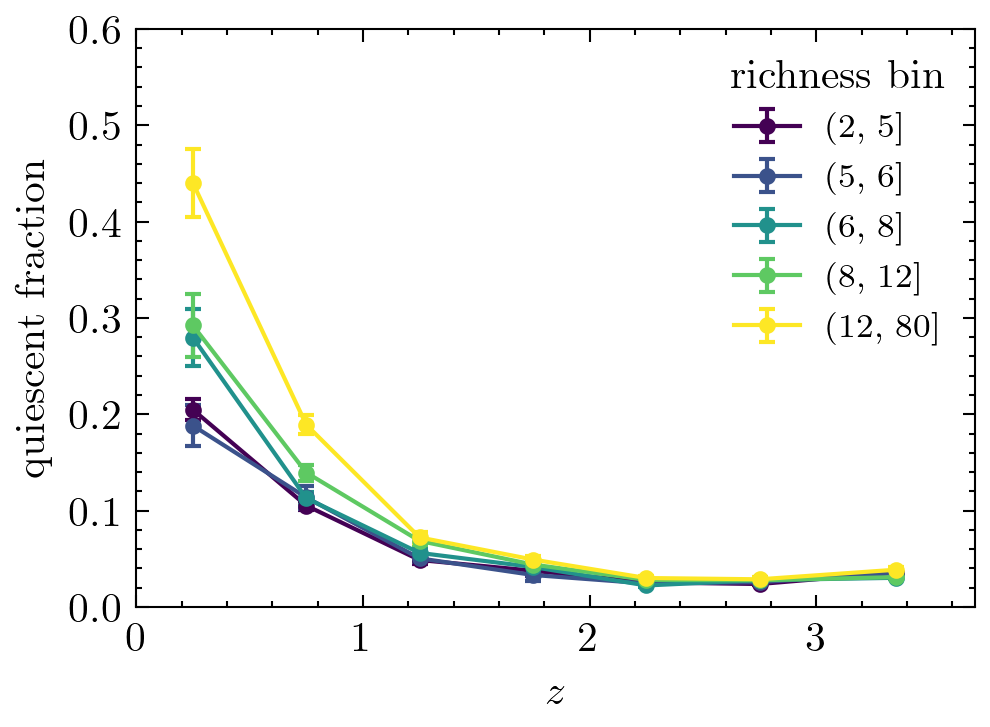}
      \caption{Same as in Fig. \ref{basic_fred_fig} but without applying purity weighting.}
         \label{basic_fred_fig_nopurity}
\end{figure}

\begin{figure}
   \centering
   \includegraphics[width=8cm]{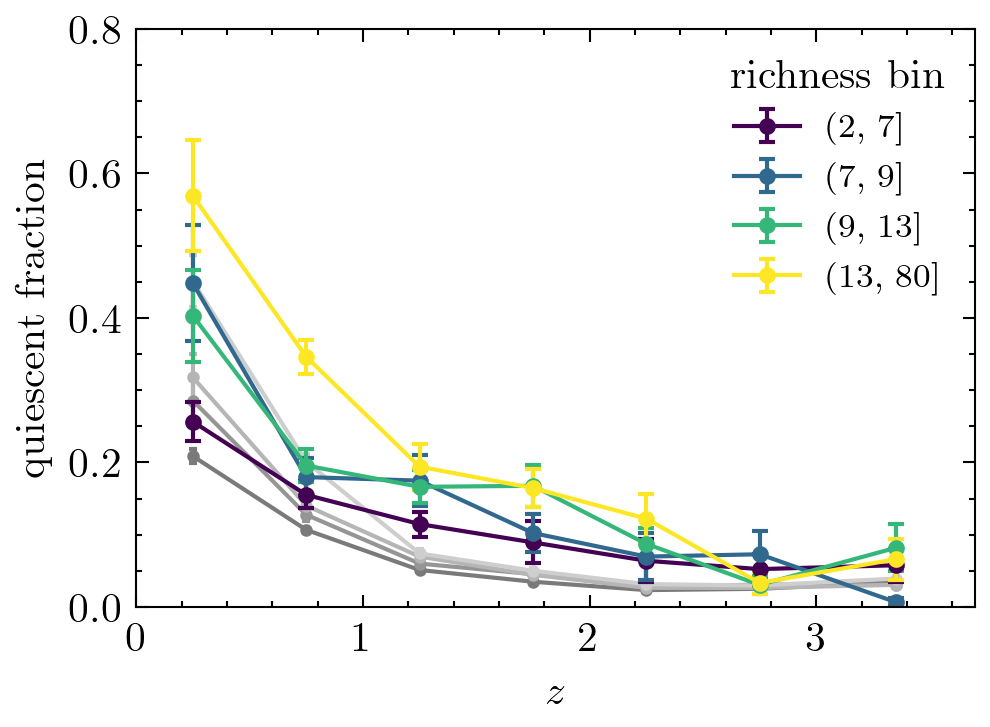}
      \caption{Same as in Fig. \ref{fred_cyl} but without applying purity weighting.}
         \label{fred_cyl_nopurity}
\end{figure}

For completeness, Figs.~\ref{basic_fred_fig_nopurity} and \ref{fred_cyl_nopurity} present the same measurements shown in Figs.~\ref{basic_fred_fig} and \ref{fred_cyl}, but without applying purity weighting in the corresponding bins for the pure-membership and cylinder methods, respectively. As expected, given the overall high purity of the COSMOS-Web catalog, all global trends are preserved. The purity weighting mainly influences the lowest-richness bins, which are naturally more prone to lower reliability and purity.

\end{appendix}

\end{document}